\documentclass{egpubl}
 
\JournalPaper         % uncomment for final version of Journal Paper
\usepackage[T1]{fontenc}
\usepackage{amsmath}
\usepackage{lmodern}
\usepackage{dfadobe}  
\usepackage{subfigure}
\usepackage[noend]{algpseudocode}
\usepackage{algorithmicx,algorithm}
\usepackage{makecell}
\biberVersion
\BibtexOrBiblatex
\usepackage[backend=bibtex,bibstyle=EG,citestyle=alphabetic,backref=true]{biblatex} 
\electronicVersion
\PrintedOrElectronic

\ifpdf \usepackage[pdftex]{graphicx} \pdfcompresslevel=9
\else \usepackage[dvips]{graphicx} \fi

\usepackage{egweblnk} 
\title{Faster Ray Tracing through Hierarchy Cut Code}

\author[WeiLai Xiang \& FengQi Liu \& Dan Li \&ZaoNan Tan \& PengZhan Xu \& MeiZhi Liu \& Qilong Kou]
{\parbox{\textwidth}{\centering WeiLai Xiang$^{1}$\orcid{0000-0002-0222-9904}, FengQi Liu$^{1}$, Dan Li$^{1}$\footnotemark,
ZaoNan Tan$^{1}$,
PengZhan Xu$^{2}$, MeiZhi Liu$^{2}$ and QiLong Kou$^{2}$
        }
        \\
{\parbox{\textwidth}{\centering $^1$ School of Computer Science and Technology, Huazhong University of Science and Technology, Wuhan, China\\
         $^2$ Tencent Technology (Shenzhen) Co., Ltd. Shenzhen, China
       }
}
}
\begin{document}

\teaser{

 \includegraphics[width=\linewidth]{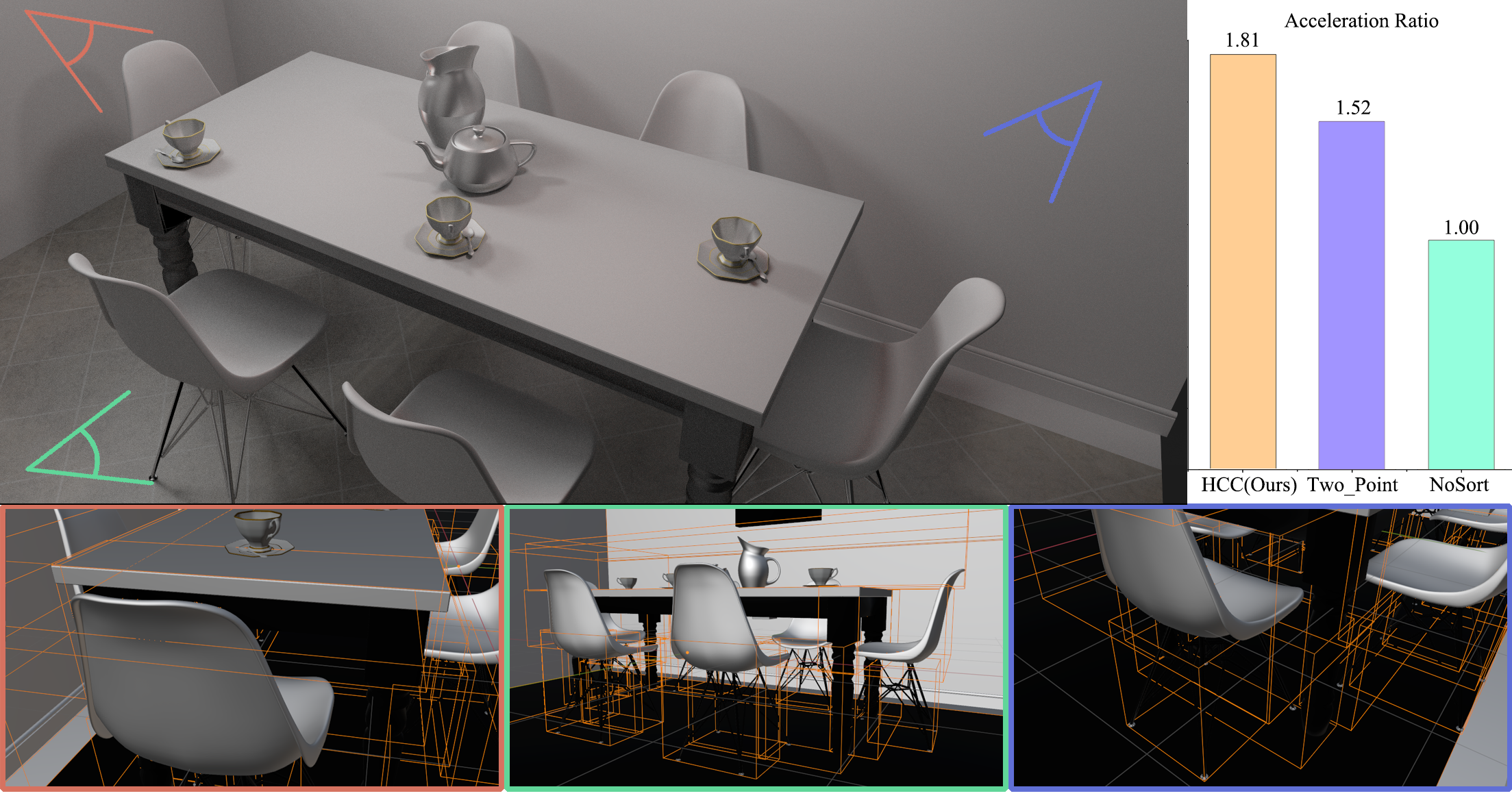}
 
  \caption{The hierarchy cut code can greatly accelerate the speed of ray tracing. We conducted a performance comparison of our reordering technique with the NoSort method and the TwoPoint reordering technique in the Breakfast scene. The scene was rendered at a resolution of 1920x1080, with two shadow rays per bounce and eight samples per pixel. The bar chart illustrating the acceleration ratio of secondary ray tracing clearly demonstrates that our method outperforms the TwoPoint method in terms of acceleration effect. The following three small figures depict the bounding boxes of each cut nod utilized in our method from three different observation positions. It is worth noting that our method employs these bounding boxes, rather than relying on spatial information, for ray encoding.}
\label{fig:teaser}
}

\maketitle
%-------------------------------------------------------------------------
\begin{abstract}
We propose a novel ray reordering technique designed to accelerate the ray tracing process by encoding and sorting rays prior to traversal. Our method, called "hierarchy cut code," involves encoding rays based on the cuts of the hierarchical acceleration structure, rather than relying solely on spatial coordinates. This approach allows for a more effective adaptation to the acceleration structure, resulting in a more reliable and efficient encoding outcome. Furthermore, our research identifies "boundary drift" as a major obstacle in achieving better acceleration effects using longer sorting keys in existing reordering methods. Fortunately, our hierarchy cut code successfully overcomes this issue, providing improved performance in ray tracing. Experimental results support the effectiveness of our approach, showing up to a 1.81 times faster secondary ray tracing compared to existing methods. These promising results highlight the potential for further enhancement in the acceleration effect of reordering techniques, warranting further exploration and research in this exciting field.

%-------------------------------------------------------------------------
%  ACM CCS 1998
%  (see https://www.acm.org/publications/computing-classification-system/1998)
% \begin{classification} % according to https://www.acm.org/publications/computing-classification-system/1998
% \CCScat{Computer Graphics}{I.3.3}{Picture/Image Generation}{Line and curve generation}
% \end{classification}
%-------------------------------------------------------------------------
%  ACM CCS 2012
%   (see https://www.acm.org/publications/class-2012)
%The tool at \url{http://dl.acm.org/ccs.cfm} can be used to generate



%\printccsdesc   
\end{abstract} 

\footnotetext{$\dagger$ Corresponding author. lidanhust@hust.edu.cn}  
%-------------------------------------------------------------------------
\section{Introduction}

Path tracing is a computationally demanding rendering technique that achieves global illumination for synthesizing realistic images through recursive ray-casting. It involves three key steps: computing the intersection between rays and the scene, shading the ray hits, and generating new rays based on the intersection results. To expedite the intersection test between rays and the scene, acceleration structures like bounding volume hierarchies are employed.

However, due to ray incoherence, parallel units (such as a warp on NVIDIA GPUs) containing rays may access disparate portions of the acceleration structure. Even if each ray only accesses a small section of the structure, the warp still needs to access numerous nodes within the acceleration structure. This results in significant bandwidth consumption, leading to a decrease in parallel efficiency.

One solution to address this issue is the implementation of ray reordering techniques. This technique involves sorting the rays prior to the tracing process, ensuring that rays computed by the same parallel unit follow similar acceleration structure traversal paths to reduce parallel divergence. To facilitate this reordering process, the rays need to be encoded as sorting keys. Existing ray encoding methods typically encode rays based on their origins and directions \cite{Aila2010, Costa2015, Reis2017}, or their origins and termination points \cite{Moon2010}.

However, there are two important challenges that need to be addressed in these spatial coordinate based reordering techniques. Firstly, the overhead, particularly for ray sorting, is relatively high. Secondly, these techniques have encountered performance bottlenecks, as their performance fails to benefit from longer encoding keys. Despite the potential for more precise ray encoding, longer encoding keys do not result in performance improvements.

Encoding rays based on spatial coordinates involves creating an implicit grid of the scene space. However, the boundaries of the cells in the grid differ from the boundaries of the nodes in the acceleration structure (Figure \ref{Fig_Boundary_Drift}). This diffrence, which we refer to as boundary drift, is the primary cause of performance bottlenecks. After conducting a comprehensive review of the relationship between rays and acceleration structures, as well as the correlation between ray spatial coherence and tracing efficiency, we have identified boundary drift as the main hindrance to achieving optimal performance.

Building on this analysis, we provide theoretical explanations for the limitations of current reordering technologies. Furthermore, through a series of experiments, we demonstrate the existence of a more efficient ray order that effectively mitigates the issues associated with boundary drift. Moreover, we introduce a novel ray encoding method called hierarchy cut code. This encoding method utilizes a cut of the acceleration structure to encode rays, effectively avoiding boundary drift and yielding improved results. The cut is determined through a priority queue search using hierarchically accumulated density as the priority metric. Comparative experiments with other methods demonstrate the superior performance of our encoding method across various scenes. Additionally, we propose a compression scheme that effectively reduces the overhead. Figure\ref{fig:teaser} presents a comparison of the secondary ray trace speeds achieved by our reordering technique, the NoSort method, and the TwoPoint reordering technique. It is evident that our method achieves significant speedup.

The main contributions of this work include:
\begin{itemize}
\item A novel encoding method based on the cut of acceleration structure to achieve better performance than existing reordering techniques.
\item An explanation of performance defects to existing reordering techniques. Especially to explain why they cannot benefit from longer sorting keys.
\item A compression scheme which can reduce the overhead by shortening the sorting key. 
\end{itemize}

%-------------------------------------------------------------------------
\section{Related Work}
One of the time-consuming steps in path tracing is the computation of ray-object intersections in the scene. To accelerate this process, researchers have explored parallel computing techniques. Some notable works in this area include the packet ray trace \cite{Gunther2007} and the ray stream trace \cite{Tsakok2009} \cite{barringer2014dynamic} . However, SIMD utilization often face challenges related to low parallel rates. Aila and Laine \cite{Aila2009} noticed this problem and discussed several possible improvements: replacing terminated rays, utilizing work queues, and designing wide trees though these methods do not pay off on the GPU of that time. So Aila and Karras \cite{Aila2010} proposed another BVH optimization algorithm based on treelet restructuring. However, the idea of replacing terminated rays inspire other reserchers. Wald \cite{Wald2011} introduced active thread compaction, which improved the efficiency of GPU path tracing by utilizing the idea of replacing terminated rays. Antwerpen \cite{Antwerpen2011} adopted a similar method to speed up Monte Carlo light transport. Laine et al. \cite{Laine2013} proposed a small kernel rendering framework that splited the intersecting and shading operations into different kernels. By minimizing the number of registers used by each thread and ensuring that terminated rays do not occupy any thread, the small kernel approach maximizes the parallel execution of threads. These advancements in parallel computing techniques and termination strategies contribute to improving the efficiency of ray-object intersection computations in path tracing algorithms.

In addition to parallel computing, hierarchical acceleration structures are commonly employed to expedite the ray tracing process. One widely used structure is the Bounding Volume Hierarchy (BVH), which has been extensively studied and optimized. Goldsmith and Salmon \cite{Goldsmith1987} introduced the first framework for constructing BVH structures, while later, MacDonald and Booth \cite{MacDonald1990} utilized the Surface Area Heuristic (SAH) to build optimized BVHs. More recently, Lin et al. \cite{Lin2020} proposed the dual-split tree as an acceleration structure, which effectively reduces the number of ray-plane intersection tests.

However, optimizing the partitioning of bounding boxes alone does not fully address the issue of high memory traffic during BVH-accelerated ray tracing. Current GPUs are highly sensitive to memory traffic, and excessive memory access can significantly impact ray tracing efficiency. One potential solution is to use wider trees. Ylitie et al. \cite{Ylitie2017} demonstrated that efficient incoherent ray traversal can be achieved through a compressed wide BVH, highlighting the importance of reducing memory consumption in the hierarchy to achieve higher performance. Similarly, Pharr et al. \cite{Pharr1997} divided the scene into cells and computed ray-cell intersections, reducing the need for disk-to-memory scene switches, especially in extremely large scenes. Building upon this idea, Navratil et al. \cite{Navratil2007} extended the approach by suggesting the splitting of both geometries and rays. Additionally, Hendrich et al. \cite{Hendrich2019} proposed a ray classification scheme that maps rays to interior nodes deeper in the tree, effectively skipping nodes in higher levels. This method helps reduce the number of traversal steps required. All these methods aim to modify the acceleration structures to reduce divergence and improve performance.

In addition to the structure of the BVH, the divergence of ray traversal paths also contributes to high memory traffic. This divergence arises from the fact that rays tend to follow different paths during BVH traversal, resulting in increased global memory access load. Meister et al. \cite{Meister2021} summarized a series of BVH methods, some of which aim to tackle this problem. Increasing the coherence of rays is a direct approach to addressing the issues associated with divergent traversal paths. One way to achieve this is by consciously generating coherent ray sets. For example, Szirmay-Kalos and Purgathofer \cite{Laszlo1999} proposed global ray bundle tracing, while Nimier-David et al. \cite{David2019} implemented coherent MCMC sampling. These methods generate highly coherent workloads, resulting in faster intersection computations. However, it is worth noting that a common drawback of these approaches is that they often require a specialized rendering pipeline.

Without altering the rules for generating rays, it is still possible to generate a coherent sequence of rays using reordering techniques. These techniques involve encoding and sorting the rays to exploit their coherence. For example, Havran and Bittner \cite{Havran2001} proposed the use of Longest Common Traversal Sequences (LCTS) to leverage coherence in ray shooting with BSP trees. Gunther et al. \cite{Gunther2007} enforced the SIMD structure to operate on coherent ray sequences, while Boulos et al. \cite{Boulos2008} explicitly reordered rays into coherent ray packets. Although ray reordering introduces additional work, Meister et al. \cite{Meister2020} demonstrated that it can significantly accelerate ray tracing speed on modern GPU architectures. Moreover, the benefits of coherent ray sequences extend to other stages of the rendering process. For instance, Eisenacher et al. \cite{Eisenacher2013} sorted termination points to improve shading performance, and Moon et al. \cite{Moon2010} employ sorting based on estimated termination points to achieve cache-coherent memory access in out-of-core rendering. Additionally, Costa et al. \cite{Costa2015} proposed a ray sorting approach based on ray direction for ambient occlusion computations.

Another approach to exploit a coherent ray sequence is through breadth-first packet traversal. Garanzha and Loop \cite{Garanzha2010} utilized this technique to reduce divergence in computations. The breadth-first packet traversal method employs a hash-based approach inspired by Arvo and Kirk's 5D ray tracing \cite{Arvo1987}. In a similar vein, Reshetov et al. \cite{Reshetov2005} proposed a method that grouped rays into frusta to minimize the number of intersection tests. Torres et al. \cite{Torres2011} further extended this idea by splitting the BVH into a forest of disjoint subtrees based on their cut, which can be seen as a form of breadth-first packet traversal in a treelet structure. While breadth-first packet traversal can accelerate primary ray and deterministic ray tracing, it may not be suitable for path tracing, as the frusta tend to become too large and intersect a significant portion of the scene. Therefore, alternative strategies are typically employed for path tracing scenarios to handle the increased complexity of intersecting rays with the scene.

%-------------------------------------------------------------------------

%-------------------------------------------------------------------------
\section{Background}

%In this section, we describe a precise definition of coherence in ray tracing. There are two kinds of coherence to be aware of: traversal path coherence and ray spatial coherence, which will be explained in Section 3.1. The word "Path" in the text refers to the traversal path of rays in the BVH. In Section 3.2, we point out that existing reordering techniques suffer from encoding bias between code boundary and node boundary. This phenomenon is called boundary drift. We propose prefix path encoding method based on traversal paths in Section 3.3. This method proves that the reordering technique still has great potential for improvement.

%-------------------------------------------------------------------------
\subsection{The Principle of Rays Reordering}
In most ray tracing systems, the ray-object intersection process is accelerated using BVH. By employing this structure, rays only need to perform intersection tests with a small subset of triangles and some BVH nodes, rather than the entire set of triangles. The BVH nodes accessed by each ray form its traversal path. However, since different rays may have different traversal paths, each parallel unit (such as a thread in the warp on NVIDIA GPU) needs to fetch multiple decentralized nodes during one intersection test. This consumes a significant amount of bandwidth and reduces the parallelization efficiency of this type of computation.

\begin{figure}[htbp]
\centering
\includegraphics[width=0.65\linewidth]{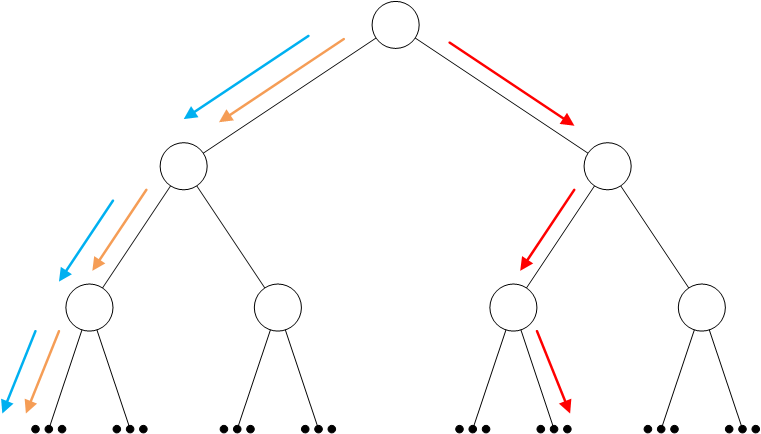}
\caption{The BVH traversal paths for rays of high traversal path coherence (orange and blue arrows) or low traversal path coherence (orange and red arrows).}
\label{Fig_traversal_path}
\end{figure}

\textbf{Traversal path coherence} is directly related to the mentioned divergence, referring to how the traversal paths of rays overlap with each other. Rays with higher traversal path coherence exhibit improved tracing efficiency, as they experience less divergence in parallel computation and access fewer nodes overall. This concept is illustrated in Figure \ref{Fig_traversal_path}, where the tree-like structure represents the initial three layers of a BVH. The arrows of three different colors depict the traversal paths of three rays within the BVH. The orange and blue arrows exhibit highly similar traversal paths, indicating a high traversal path coherence for their respective rays. Conversely, the orange and red arrows show vastly different traversal paths, indicating a very low traversal path coherence for their corresponding rays.

The measurement of traversal path coherence poses a challenge as the paths cannot be known in advance before intersection tests. Instead, \textbf{ray spatial coherence} provides a more accessible measure. This concept implies that rays are closely positioned in world space, sharing similar origins and directions. Rays with similar origins and directions are more likely to intersect with the same BVH nodes and exhibit similar access patterns. Hence, rays with high ray spatial coherence tend to demonstrate high traversal path coherence. Similar conclusions have been presented by Moon et al. \cite{Moon2010}. 

Ray reordering technique increases the coherence of rays. The key focus of ray reordering methods lies in the computation of ray sorting keys. As the analysis before, the chosen encoding method should ensure consistency between ray spatial coherence and traversal path coherence. Traditional ray encoding methods regard rays in three-dimensional space as points in a five-dimensional space (ray space), with three dimensions representing the origin and two dimensions representing the direction. The process of ray encoding involves transforming rays into one-dimensional sorting keys such as Morton codes, which are then used to sort the rays.

%-------------------------------------------------------------------------
\subsection{Sorting Key Computation Methods}

\begin{figure}[htbp]
\centering
\includegraphics[width=0.65\linewidth]{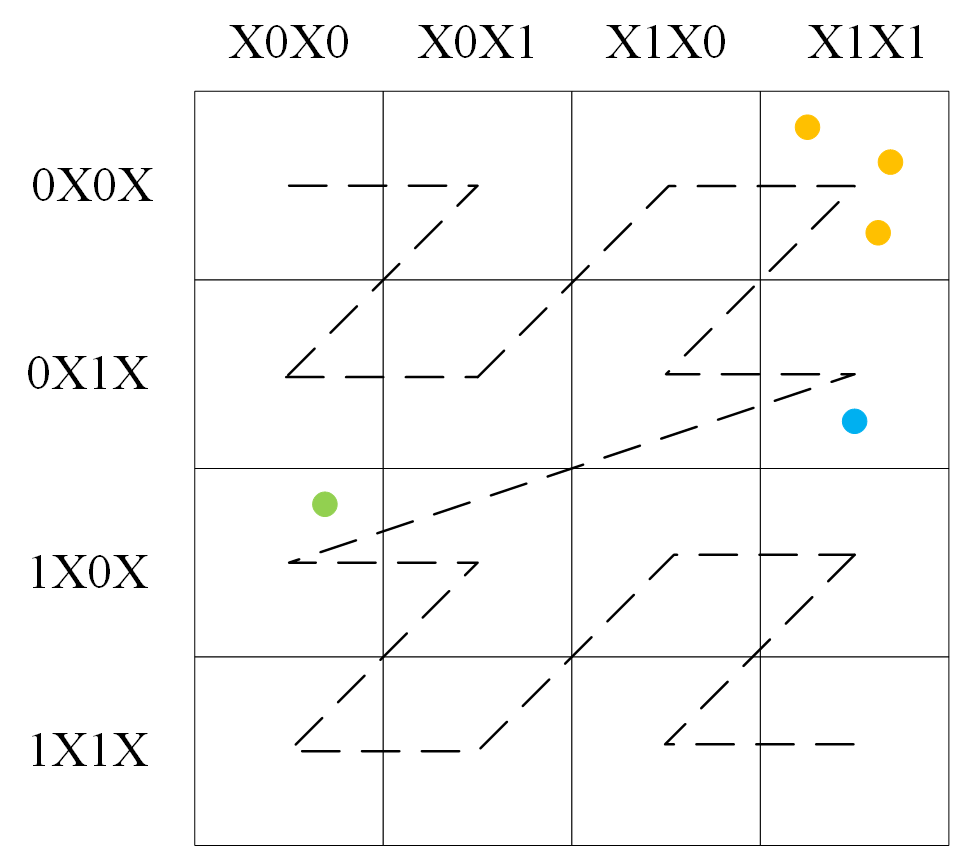}
\caption{Code cell for 4 bits 2D Morton code. Points in the same cell will have the same code. For example, three orange points in the upper left cell have the same Morton code 0101.}
\label{Fig_CB}
\end{figure}

As mentioned earlier, conventional reordering techniques rely on ray spatial coherence. They employ encoding methods to encode the rays and then sort them based on their spatial coordinates, resulting in a ray set with high ray spatial coherence. One commonly used encoding method is Morton code (Figure \ref{Fig_CB}). Each encoding method has its own code cells, meaning that points or rays within the same code cell will have the same code value under that particular encoding method.

For instance, Figure \ref{Fig_CB} illustrates the code cells of a 4-bit 2D Morton code. The dashed line represents the Z-curve, which is a visualization of the encoding process. The 4-bit 2D Morton code divides the space into a grid with sixteen cells, and points within each cell will share the same code. In the example, the three orange points in the upper right corner will all be encoded as 0101. The blue point will have the code 0111, and the green point will have the code 1000. Consequently, the sixteen solid black cells represent the code cells for this specific encoding.

Several techniques based on space-filling curves have already been proposed. We briefly summarize here. More details can be found in \cite{Meister2020}.

\textit{Origin}. This method utilizes the origins of rays as a sorting key to create a 3D space-filling curve.

\textit{Origin-Direction}. This method was proposed by Reis et al. \cite{Reis2017} for sorting secondary rays when constructing a ray space hierarchy. They include the direction into the sorting key using lower bits of the key for the parametrized direction to improve ray coherence.

\textit{Direction-Origin}. This method sets the direction information in higher bits of the code followed by the bits representing the origin. It was proposed by Costa et al. \cite{Costa2015} for sorting shadow and ambient oc-clusion rays.

\textit{Origin-Direction Interleaved}. Another strategy of computing sorting keys uses interleaving of bits representing origin and direction. This technique was used by Aila and Karras \cite{Aila2010} to study the behavior of tracing incoherent rays. 

\textit{Two Point}. This method estimates the termination point and computes the sorting key using the 6D point which represents the ray. The key is constructed by interleaving bits of origin and termination points. \cite{Moon2010,Meister2020}

\begin{figure}[htbp]
\centering
\subfigure[Boundary drift in Morton code]
{
\begin{minipage}{0.68\linewidth}
\centering
\includegraphics[width=\linewidth]{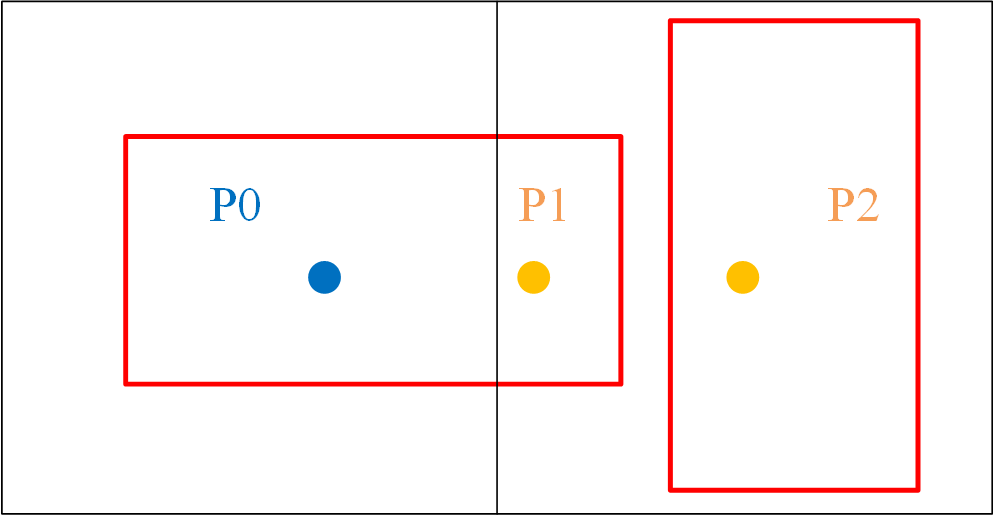}
\end{minipage}
}
\subfigure[Boundary drift in spherical coordinates code]
{
\begin{minipage}{0.67\linewidth}
\centering
\includegraphics[width=\linewidth]{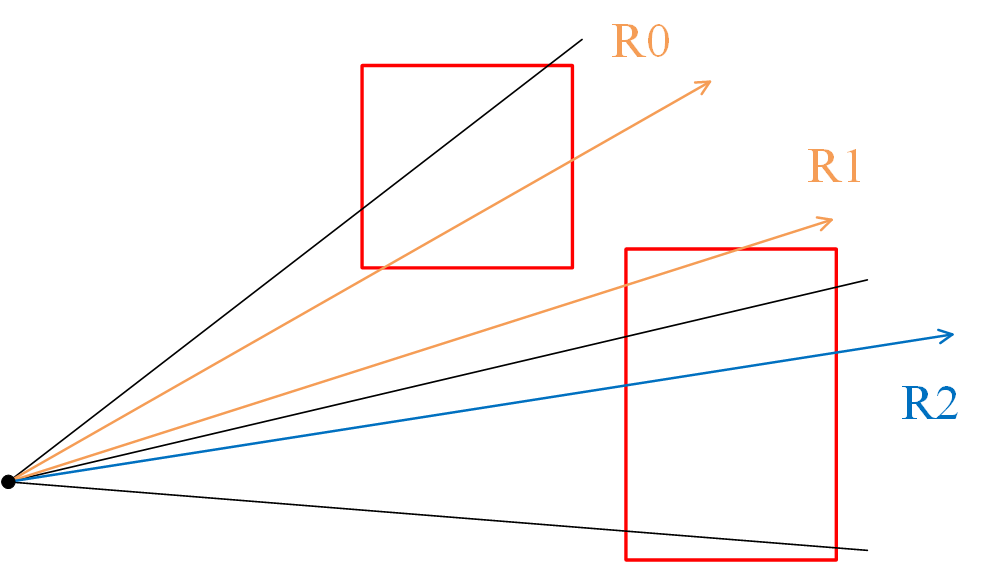}
\end{minipage}
}
\caption{Boundary drift in (a): Morton code, (b): Spherical coordinates code. The red boxes are BVH nodes' boundaries, and the black boundaries are the code cell/cone boundary. This example shows that rays in the same code boundary can intersect different BVH nodes, which causes boundary drift.}

\label{Fig_Boundary_Drift}
\end{figure}

%-------------------------------------------------------------------------
\subsection{Boundary Drift in General Reordering Techniques}

All of the aforementioned methods suffer from \textbf{boundary drift}. Boundary drift refers to the situation where the boundaries of the code cells differ from the boundaries of the BVH nodes. This can lead to inconsistent intersection results between rays that have the same code but intersect different BVH nodes. In Figure \ref{Fig_Boundary_Drift} (a), the black boundaries represent the Morton code boundaries, while the red boundaries represent the BVH node boundaries. Points P0 and P1, which fall within the same BVH node boundary, correspond to different codes because they belong to different code boundaries. Conversely, points P1 and P2, which fall within different BVH node boundaries, correspond to the same codes. The term "drift" arises from the observation that the boundary of code cells appears to be shifting away from the BVH boundary in Figure \ref{Fig_Boundary_Drift} (a). Direction encoding methods also suffer from a similar issue, as depicted in Figure \ref{Fig_Boundary_Drift} (b).

Boundary drift can have a particularly significant impact in certain scenarios, such as when complex small objects are placed within a large scene. In such cases, the hit points on those small objects may be encoded with only a few bits, leading to severe performance degradation. In fact, the performance may even be worse than when no optimization is applied at all.

%------------------------------------------------------------------------

%------------------------------------------------------------------------

\section{Reorder the Rays with Traversal Path Based Code }

Boundary drift introduces inconsistencies between the measured ray spatial coherence and traversal path coherence, leading to the realization that high ray spatial coherence does not guarantee high traversal path coherence. As a result, reordering techniques may fail to achieve optimal results in certain situations. In such cases, using sorting keys that directly represent the traversal path is a better choice. By incorporating traversal path information into the sorting keys, the reordering algorithm can better align rays with similar traversal paths, resulting in improved performance and efficiency. We propose three traversal path-based methods for computing sorting codes. The first method is the prefix path code, which completely avoids boundary drift but incurs a large overhead. Next, we introduce the hierarchy cut code and its improved version, the multi-level hierarchy cut code, which strike a balance between overhead and traversal path coherence.  

%------------------------------------------------------------------------
\subsection{Prefix Path Code}

If the encoding overhead is not taken into consideration, directly encoding the full BVH traversal path can avoid boundary drift and achieve the best traversal path coherence. However, this method has limited practical optimization value due to its large overhead. Nonetheless, it can still serve as a reasonable upper bound for performance comparison.

\begin{figure}[htbp]
\centering
\includegraphics[width=0.71\linewidth]{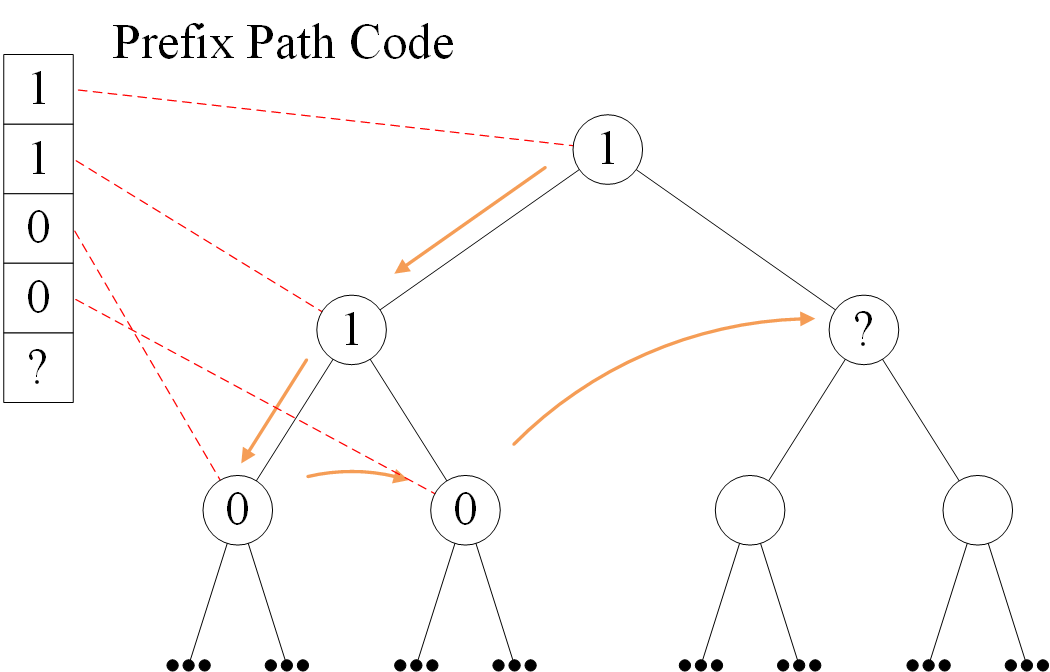}
\caption{Example of prefix path code. The orange arrows represent the traversal path, and the prefix path code is used to record each intersection result along this path. In the example, a "1" is assigned for intersections, indicating that the ray intersects with the corresponding node, while a "0" is assigned for non-intersections. For the given example path, the prefix path code is 1100, indicating that the first two intersections are true (intersected), followed by two false (not intersected) intersections.}
\label{Fig_ppc}
\end{figure}

The complete traversal path is typically long and requires a significant number of bits for encoding. To make this encoding method feasible and enable better comparison with other methods, it is necessary to limit the length of the encoding. This can be achieved by terminating the traversal after a certain number of bits have been set. The proposed encoding method is called the \textbf{prefix path code}, as it encodes the ray based on its traversal path prefix. Figure \ref{Fig_ppc} provides an illustration of this encoding method. The traversal path (represented by orange arrows) is recorded, and the prefix path code is formed by setting each bit to 1 if the corresponding node intersection result is true, and 0 otherwise. For example, in Figure \ref{Fig_ppc}, the first four intersection results of the rays are true, true, false, and false, resulting in a 4-bit prefix code of 1100. This prefix path code represents the corresponding traversal path through the hierarchy when the rays are forced to traverse the BVH in the same order.

The prefix path code consistently achieves the best ray trace speed performance in most scenes when compared to other reordering techniques of the same bit length. This finding provides further evidence for our hypothesis that traversal path coherence is highly correlated with tracing efficiency, and it underscores the potential for further performance improvements in current ray reordering techniques. Eliminating the issue of boundary drift from the encoding method can greatly enhance the practicality of reordering techniques. However, it is important to note that the overhead associated with the prefix path code is nearly equivalent to that of the actual intersection test. The significant overhead associated with prefix path code makes it impractical for implementation as an optimization scheme. However, it can still serve as a useful reference when evaluating the ray trace speed of certain methods.

%------------------------------------------------------------------------

%------------------------------------------------------------------------
\subsection{Hierarchy Cut Code} \label{section:HCC}
\subsubsection{Defination}

\begin{figure}[!htbp]
\centering
	\begin{minipage}{0.81\linewidth}
		\centering
		\includegraphics[width=\linewidth]{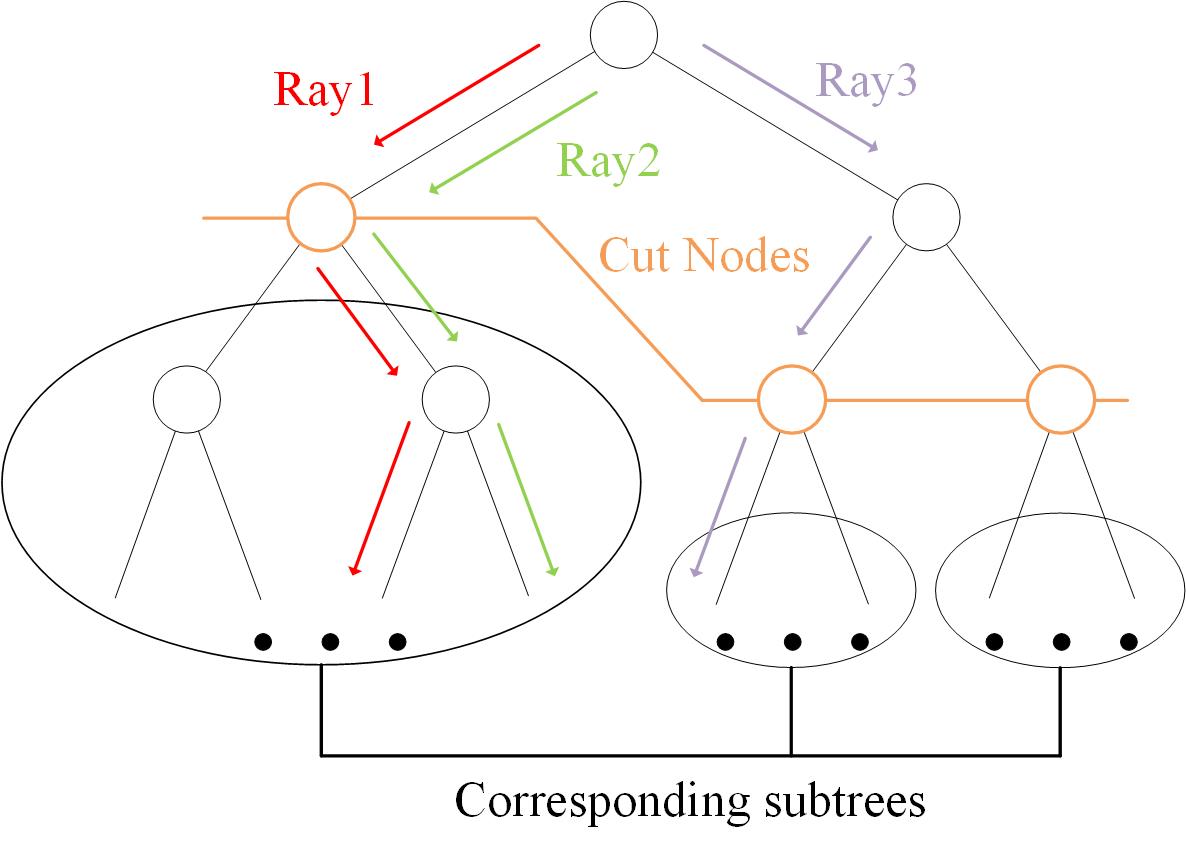}
		%\caption{}
	\end{minipage}

\caption{Example shows how hierarchy cut code is related with traversal path coherence. The example cut (highlighted in orange) in this BVH consists of three cut nodes and three corresponding subtrees. We have three example rays traversing this BVH. Ray1 and Ray2 both access the same cut node, indicating that they will traverse the same corresponding subtree. On the other hand, Ray3 does not access that particular cut node. As a result, Ray1 and Ray2 exhibit higher traversal path coherence, as they have a longer overlap in their traversal paths.}
\label{Fig_HCC}
\end{figure}

The primary objective of improving the encoding method is to eliminate boundary drift. In order to achieve this goal without introducing significant overhead, we utilize a cut of the BVH to represent the entire BVH. Instead of encoding the rays based on their traversal path, this method encodes the rays based on their intersection results with the chosen cut. This intersection result of the cut nodes can implicitly indicate the traversal path. We refer to this encoding as "hierarchy cut code" (HCC). Figure \ref{Fig_HCC} illustrates how this code relates to traversal path coherence.

When a ray intersects a cut node, it will traverse the corresponding subtree. Rays that traverse the same subtree are more likely to have overlapping traversal paths. The degree of overlap in traversal paths can be measured by the bit difference between their hierarchy cut codes. The more identical the sequence of subtrees accessed by two rays, the higher the likelihood of them having overlapping traversal paths. Therefore, traversal path coherence can be quantified by the bit difference between their hierarchy cut codes. The more cut nodes two rays access in common, the higher their traversal path coherence.

The hierarchy cut code requires only the calculation of intersection results between rays and fixed cut nodes. The encoding process is outlined in Algorithm \ref{Alg:1}. The cut nodes are shared among all rays, allowing them to be fetched into shared memory before the ray intersection test (Line 1). For each ray, the intersection result with each cut node is computed and recorded as the hierarchy cut code (Lines 5 and 6). The bits order is determined by cut nodes order. We explain it in \ref{section:HAD}.

\begin{algorithm}[h]
\caption{The process of hierarchy cut encoding}
\begin{algorithmic}[1]
\State copy cut nodes from global memory into shared memory;
\While{have ray to encode}
\State Fetch $Ray$ from global ray queue;
\ForAll{$Cut Node$ in cut nodes}
\State Compute intersection between $Cut Node$ and $Ray$;
\State Set corresponding bits in HCC;
\EndFor
\EndWhile
\end{algorithmic}
\label{Alg:1}
\end{algorithm}

%------------------------------------------------------------------------
\subsubsection{Determining Cut Nodes} \label{section:HAD}

\begin{figure}[!htbp]
\centering
\subfigcapskip=3pt 
\subfigure[An example BVH.]{
\begin{minipage}{0.3\linewidth}
\centering
\includegraphics[width=\linewidth]{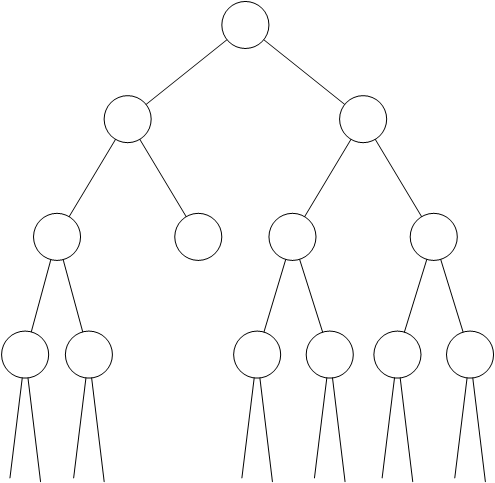}
\end{minipage}
}
\subfigure[Cut node search starts from root node.]{
\begin{minipage}{0.3\linewidth}
\centering
\includegraphics[width=\linewidth]{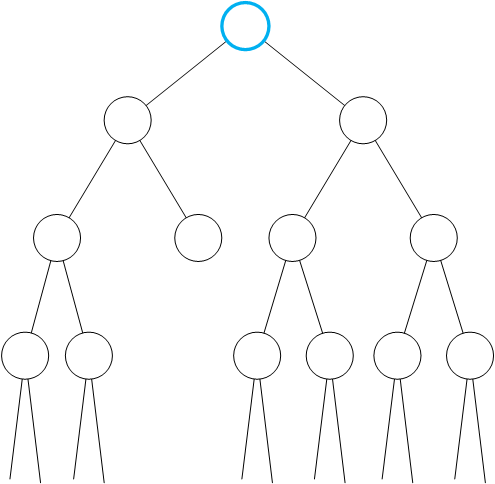}
\end{minipage}
}

\subfigure[Breadth-first search to determine cut nodes.]{
\begin{minipage}{0.3\linewidth}
\centering
\includegraphics[width=\linewidth]{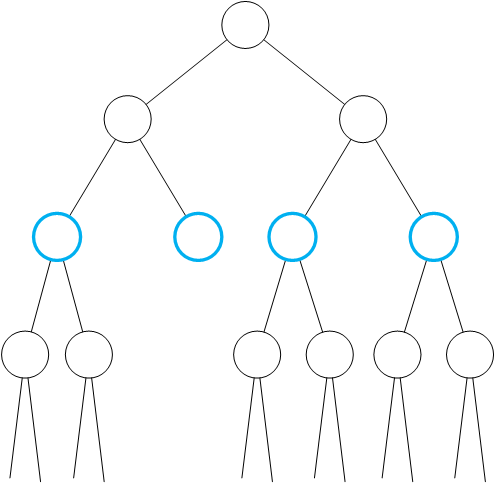}
\end{minipage}
}
\subfigure[Special case when encount leaf node.]{
\begin{minipage}{0.3\linewidth}
\centering
\includegraphics[width=\linewidth]{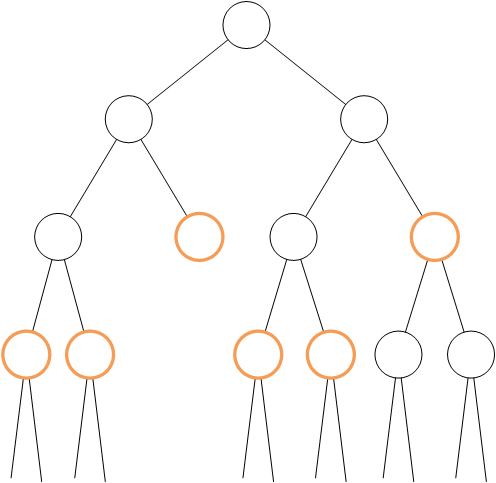}
\end{minipage}
}
\caption{Demonstration of how to determine cut nodes. (a) This is an unbalanced BVH. (b) The search starts from the root node of BVH. The blue node is in the search queue. (c) The search goes in breadth-first order. Four nodes are in the search queue at this moment, one leaf node and three top nodes. (d) Stop searching when enough cut nodes are found. Seven cut nodes (colored in orange) are found in this case. We treat the leaf node as cut node and no longer search this branch.}
\label{Fig_BF}
\end{figure}

The cut nodes remain static throughout the entire process. Figure \ref{Fig_BF} illustrates the search process. The process begins with a priority queue that initially contains the root node of the tree. The search process removes the head node from the queue and adds its child nodes to the queue until the desired number of nodes is found. Any nodes remaining in the queue after the search process are considered cut nodes. Typically, the cut nodes will not be leaf nodes. However, in some cases of highly unbalanced BVH, a leaf node may be chosen as a cut node. Figure \ref{Fig_BF} (d) illustrates such a scenario. Since leaf nodes cannot be divided further, they are treated as cut nodes and no longer participate in the search process. This measure ensures that the search process does not encounter failure when encountering leaf nodes. The pseudo code for the search process is provided in Algorithm \ref{Alg:2}.

\begin{algorithm}[h]
\caption{Compute the cut nodes}
\begin{algorithmic}[1]
\State Clear the $CutNodeList$;
\State Set $MaxQueueSize$ to required length;
\State Copy the root node into $PriorityQueue$;
\While{len($PriorityQueue$) < $MaxQueueSize$}
\State Pop $Node$ from $PriorityQueue$;
\If{$Node$ is leaf}
\State Add $Node$ into $CutNodeList$;
\State $MaxQueueSize$ -= 1
\Else
\State Push the child of $Node$ into $PriorityQueue$;
\EndIf
\EndWhile
\State Add all $Node$ in $PriorityQueue$ into $CutNodeList$;
\end{algorithmic}
\label{Alg:2}
\end{algorithm}

The priority of nodes is a significant aspect to consider in the generation of a uniform cut. We define the density of leaf nodes as the number of triangles they contain, while the density of non-leaf nodes is determined as the sum of the densities of their child nodes, along with their surface area. This priority function is referred to as "hierarchically accumulated density" (HAD). By utilizing HAD, we can automatically allocate more bits to areas with denser meshes. The following Formula \ref{eq1} shows the detailed definition of HAD. TC represents the number of triangles associated with the leaf node, SA represents the surface area of node n, and the subscripts "left" and "right" denote the left and right children of node n, respectively. The hyperparameter $\lambda$ is used to balance the weight between internal nodes and leaf nodes, and in our experiments, we set its value to 0.125.

\begin{small}
\begin{equation}
	HAD_n = \begin{cases}
		TC_n * SA_n&\mbox{n is leaf node} \\ 
		HAD_{left} + HAD_{right} + \lambda*SA_n&\mbox{n is internal node}
	\end{cases}
	\label{eq1}
\end{equation}
\end{small}

It is important to distinguish between HAD and the Surface Area Heuristic (SAH). While both methods consider the surface area of nodes, their underlying purposes and approaches differ. HAD goes beyond SAH by also incorporating subtree information corresponding to nodes. On the other hand, SAH focuses solely on the surface area of nodes and is primarily used to optimize the construction of the bounding volume hierarchy (BVH) in a top-down process. Another distinction is the timing of computation. SAH is typically computed during the construction of the BVH, as it guides the decision-making process for dividing nodes. In contrast, HAD is computed after the BVH construction, as it requires subtree information for each internal node. This makes HAD a bottom-up process that analyzes the already constructed hierarchy.

Cut nodes are sorted in descending order based on their HAD values. Nodes with higher HAD values, indicating larger surface areas or bigger subtrees, have a greater impact on the traversal path coherence of rays. Therefore, the intersection results with these nodes are assigned higher bits in the HCC codes, making them more significant during the subsequent sorting process. By assigning higher bits to intersection results with nodes of higher HAD values, the sorting algorithm prioritizes these results and ensures that rays with similar traversal paths are grouped together.     
%------------------------------------------------------------------------

%------------------------------------------------------------------------
\subsubsection{Other Details about Hierarchy Cut Code}

\begin{figure}[!htbp]
\centering
%\subfigcapskip=3pt 
%\subfigure[Four rays encoded by hierarchy cut code.]{
%	\begin{minipage}{0.83\linewidth}
%		\centering
%		\includegraphics[width=0.83\linewidth]{IMG/Figure5-Newa}
		%\caption{}
%	\end{minipage}
%}
%\subfigure[The code boundary of hierarchy cut code.]{
	
%	\begin{minipage}{0.83\linewidth}
		%\centering
		\includegraphics[width=0.56\linewidth]{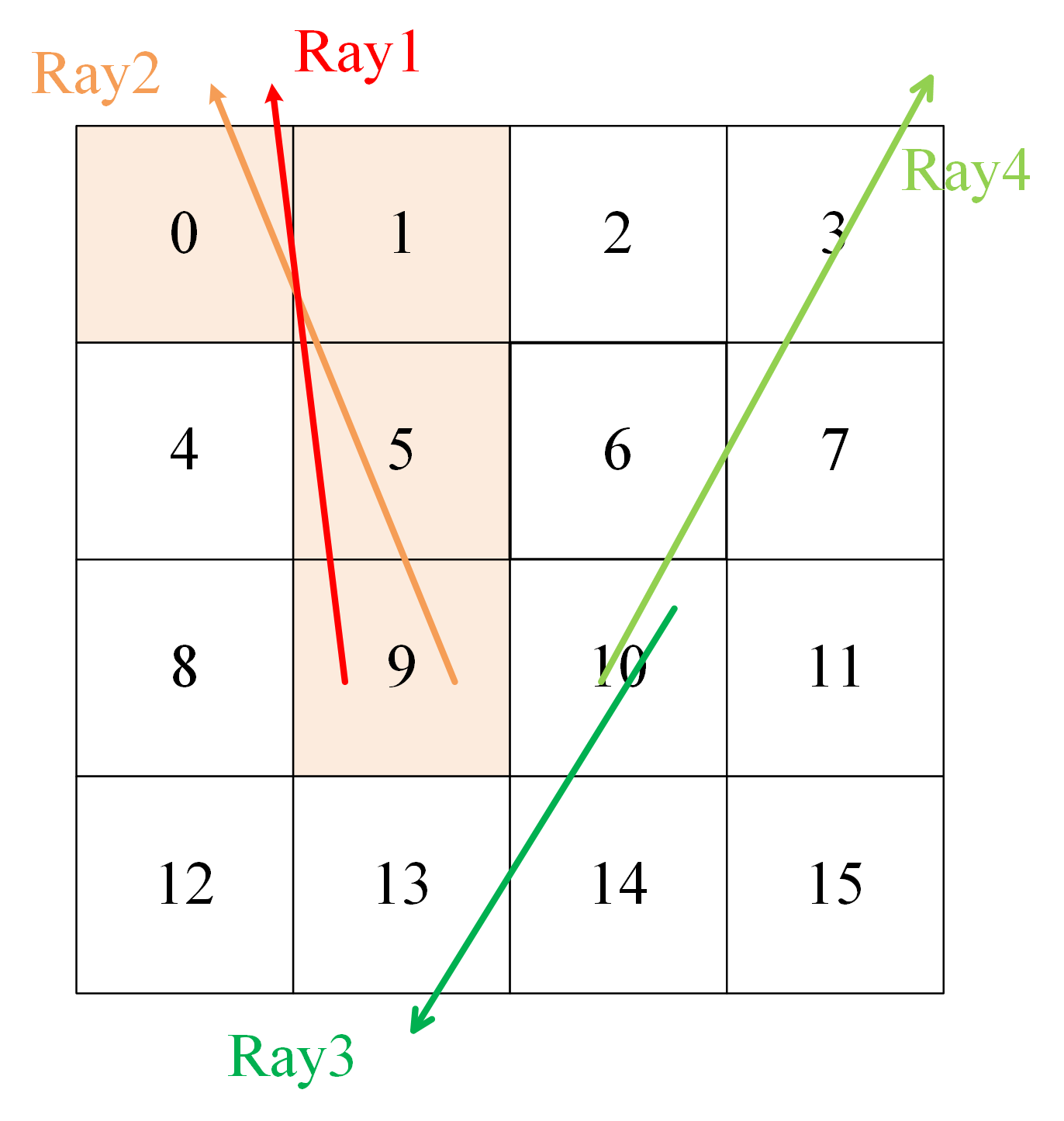}
		%\caption{}
%	\end{minipage}
%}
\caption{Demonstration of how hierarchy cut code encodes rays' spatial information. The sixteen cells of the grid are the cut nodes' boundaries. Four rays are encoded by sixteen cut nodes. Ray1 and Ray2 are close to each other. Ray3 and Ray4 have reverse direction and close origin. Ray1 traverses through nodes 0, 1, 5, and 9 while bypassing other nodes. Consequently, the first, second, sixth, and tenth bits of Ray1's HCC are set to 1, while the remaining bits are set to 0. Hence, the HCC of Ray1 is 1100010001000000. Other rays corresponding codes are: Ray2-1100010001000000, Ray3-0000000000101110, Ray4-0001001100100110.}
\label{Fig_SpaceAnchor}
\end{figure}

\textit{Sorting}. Hierarchy cut codes do not have a partial order. In fact, HCC is more like a classification label. As shown in Figure \ref{Fig_SpaceAnchor}, cut nodes divide the whole scene into several areas. The HCC of a ray marks which areas are passed by the ray and which are not, so the difference between rays is measured by the bit difference of their HCC. Rays with similar HCC (that is, they pass through similar areas) should be clustered together to increase their path coherence. Sorting the HCCs as natural numbers is a good method to achieve this goal, although in the sorting result, some rays that are close in order may have a high bit difference, such as 1000 and 0111 (indicating they pass through totally different areas). The experiment shows that our encoding method is relatively rough, resulting in most warps processing rays with the same hierarchy cut code. If the rays in a warp have the same code, the order of the code does not matter. Ultimately, it is the traversal path coherence within the warps that ultimately affects the tracing efficiency.

\textit{Ray direction}. Hierarchy cut code can achieve good performance without explicitly encoding ray direction. In most cases, hierarchy cut code can implicitly represent ray direction. As illustrated by Ray3 and Ray4 in Figure \ref{Fig_SpaceAnchor}, if a ray's origin is within the scene, reversing its direction may result in the ray accessing completely different cut nodes. Only when two rays have reverse start and end points on the scene boundary, their corresponding hierarchy cut codes will be duplicated. Additionally, adding direction information to the code would increase the risk of boundary drift. Experiments on this phenomenon are included in the appendix.

\subsection{Multi-Level Hierarchy Cut Code}\label{MLHCCs}

One drawback of our hierarchy cut code is the limited space division. The 32-bit hierarchy cut code divides the space into 32 blocks in total, which may not provide sufficient granularity. In comparison, conventional encoding methods using 32 bits (16 bits for origin and 16 bits for direction) can achieve a much finer division of space with 32*32*64 blocks (5 bits for x, 5 bits for y, and 6 bits for z). To address this limitation, we introduce multi-level hierarchy cut code (MLHCC), which utilizes multi-level cut nodes.

\begin{figure}[!htbp]
\centering
\subfigcapskip=3pt 
\includegraphics[width=0.58\linewidth]{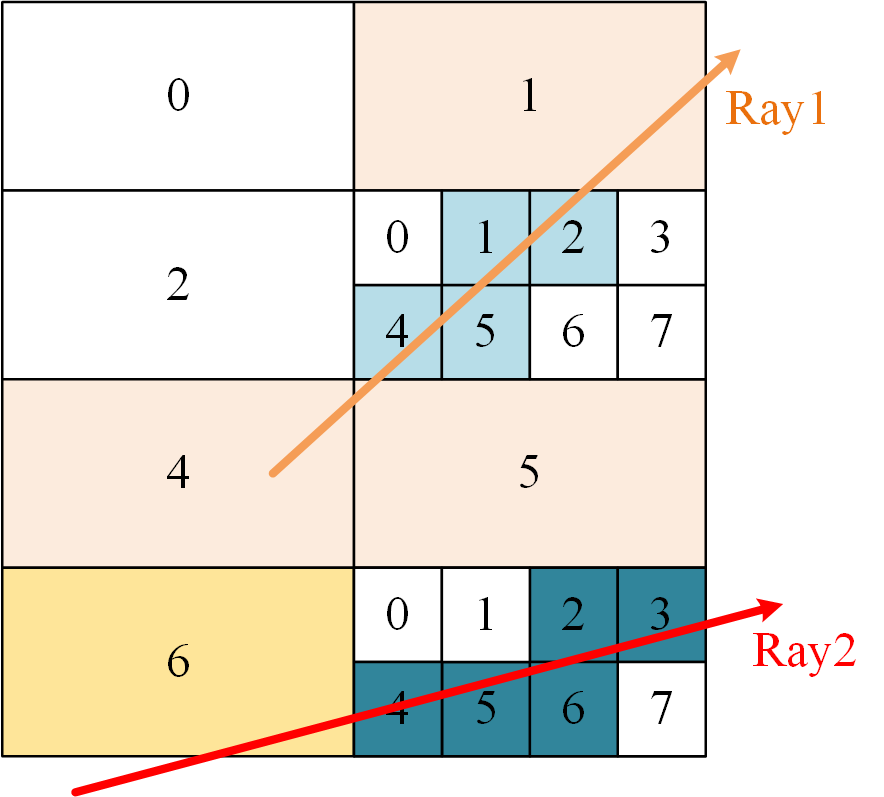}
\caption{Demonstration of how multi-level hierarchy cut code encodes rays. Ray1 intersects four top level cut nodes (orange cells). Select top level cut node 3 as the representative top level cut node, Ray1 intersects four corresponding low level cut nodes (blue grids). So Ray1 can be encoded as 0101110001101100 and Ray2 can be encoded as 0000001100111110. The first 8 bits are the intersection results with top level cut nodes, and the last 8 bits are the intersection results with corresponding low level cut nodes. In this way, we can have at most 64 low level cut nodes with 16 bits, which means the space is divided into finer blocks than using 16 bits hierarchy cut code.}
\label{Fig_MLHCC}
\end{figure}

Multi-level cut nodes consist of top-level and low-level cut nodes. The top-level cut nodes divide the BVH, while the low-level cut nodes further divide the subtree corresponding to each top-level cut node. When encoding a given ray, we record the intersection results of that ray with all top-level cut nodes. Then, we select a representative top-level cut node that is passed by the ray and record the intersection results of the corresponding low-level cut nodes. This approach allows us to achieve a finer space division by utilizing more low-level cut nodes than the available bit length. Figure \ref{Fig_MLHCC} illustrates how this encoding method can create a more refined space division.

To implement MLHCC with a two-level structure, there are two challenges to address: generating the two levels of cut nodes and determining the representative top-level cut node. The generation of two-level cut nodes follows a similar method as HCC (Algorithm \ref{Alg:3}). However, when generating the two-level cut nodes, we discard the top-level node that does not have a sufficient subtree (as shown in line 6 of Algorithm \ref{Alg:3}). This ensures that there are no empty low-level nodes.

The determination of the representative top-level cut node is based on the intersection results of the ray with all top-level cut nodes. We select the top-level cut node that is passed by the ray and has the highest HAD value as the representative top-level cut node. In practice, we can sort the top-level cut nodes according to their HAD values before sending them to the GPU memory. The representative top-level cut node is then simply the first intersected top-level cut node in the sorted list.

For example, in the case of a 16-bit MLHCC, there are 8 top-level nodes and 8*8 low-level nodes (each top-level node corresponds to 8 low-level nodes) stored in GPU memory. As illustrated in Figure \ref{Fig_MLHCC}, the upper 8 bits of 16-bit MLHCC represent the HCC generated by the 8 top-level nodes, while the lower 8 bits represent the HCC generated by the 8 low-level nodes. These low-level nodes correspond to the top-level node with the highest HAD value among the top-level nodes that the ray has passed through.

\begin{algorithm}[h]
\caption{Compute the cut nodes for multi-level hierarchy cut code}
\begin{algorithmic}[1]
\State Clear the $CutNodeList$;
\State Set $MaxQueueSize$ to required length;
\State Copy the root node into $PriorityQueue$;
\While{len($PriorityQueue$) < $MaxQueueSize$}
\State Pop $Node$ from $PriorityQueue$;
\If{$Node$ has lagre enough subtree}
\State Push the child of $Node$ into $PriorityQueue$;
\EndIf
\EndWhile
\State Add all $Node$ in $PriorityQueue$ into $CutNodeList$;
\end{algorithmic}
\label{Alg:3}
\end{algorithm}

\begin{figure*}[htb]
\centering
\includegraphics[width=\linewidth]{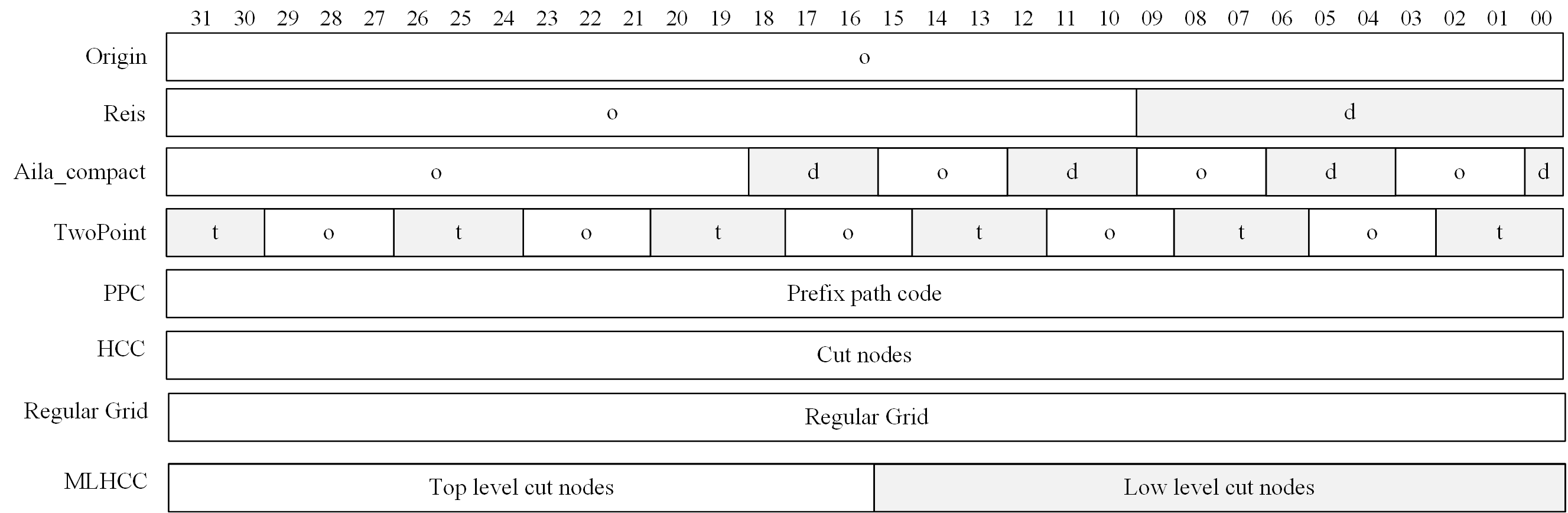}
\caption{The overview of sorting key occupied by chosen encoding methods. The figure shows bits of a 32-bit key occupied by encoding origin (o), direction(d), estimated termination point(t), or the codes introduced earlier.}

\label{Fig_coding}
\end{figure*}

The main difference between MLHCC and HCC lies in the final cut nodes. In HCC, the cut nodes are generated before ray tracing, so all the rays are encoded using the same cut nodes. On the other hand, MLHCC selects the node with the highest HAD value from the top-level cut nodes that are passed by the ray and generates low-level cut nodes in the subtree corresponding to this node. Since different rays may pass through different top-level cut nodes and generate different low-level cut nodes, the final cut nodes that determine the codes will vary among rays. This dynamic selection of cut nodes allows MLHCC to adapt to the specific ray-tracing needs of different rays.

MLHCC can be seen as a compression scheme to HCC. For instance, the 16-bit MLHCC in Figure \ref{Fig_MLHCC} can be considered a compressed form of the 64-bit HCC. However, there are two main reasons why MLHCC may not perform as well as expected when the encoding length is sufficient. The first reason is the premature discard of other top-level cut nodes' information in MLHCC. The selection of the representative top-level cut node plays a crucial role in the performance of MLHCC. If the chosen representative top-level cut node is not optimal, it can significantly impact the overall performance of MLHCC. This limitation is also the reason why MLHCC is not designed with more levels. The second reason is that while finer code can improve encoding accuracy, they do not always lead to improved tracing performance. Simply increasing the encoding accuracy by using longer sort keys does not guarantee better tracing efficiency. In Section 6.6, we will provide further explanation as to why increasing the encoding accuracy alone may not always result in improved tracing efficiency.

%------------------------------------------------------------------------

%------------------------------------------------------------------------

\section{Implementation}

To compare the performance of our technology with existing methods, we select four representative methods: Origin, Reis, Aila\_Compact, and TwoPoint, as introduced by Meister et al. \cite{Meister2020}. Among the various termination point estimation methods in the TwoPoint method, we chose the fixed-length estimation method, which sets the ray length to 0.25 of the largest extent of the scene bounding box. Other estimation methods relying on records from the last frame are not suitable for comparison.

These selected methods will be compared with our three proposed methods: hierarchy cut code (HCC), multi-level hierarchy cut code (MLHCC), and prefix path code (PPC). Additionally, we designe a \textbf{regular grid encoding} method to compare with the hierarchy cut code. The main difference between the hierarchy cut code and regular grid encoding is that the BVH cut nodes are replaced by grids that divide the space evenly. By comparing the performance of these two encoding methods, we can highlight the effect of avoiding boundary drift. An overview of all the aforementioned ray encoding methods is depicted in Figure \ref{Fig_coding}. This comprehensive comparison will provide insights into the performance and effectiveness of each method.

\begin{figure}[htbp]

\centering
\includegraphics[width=0.5\linewidth]{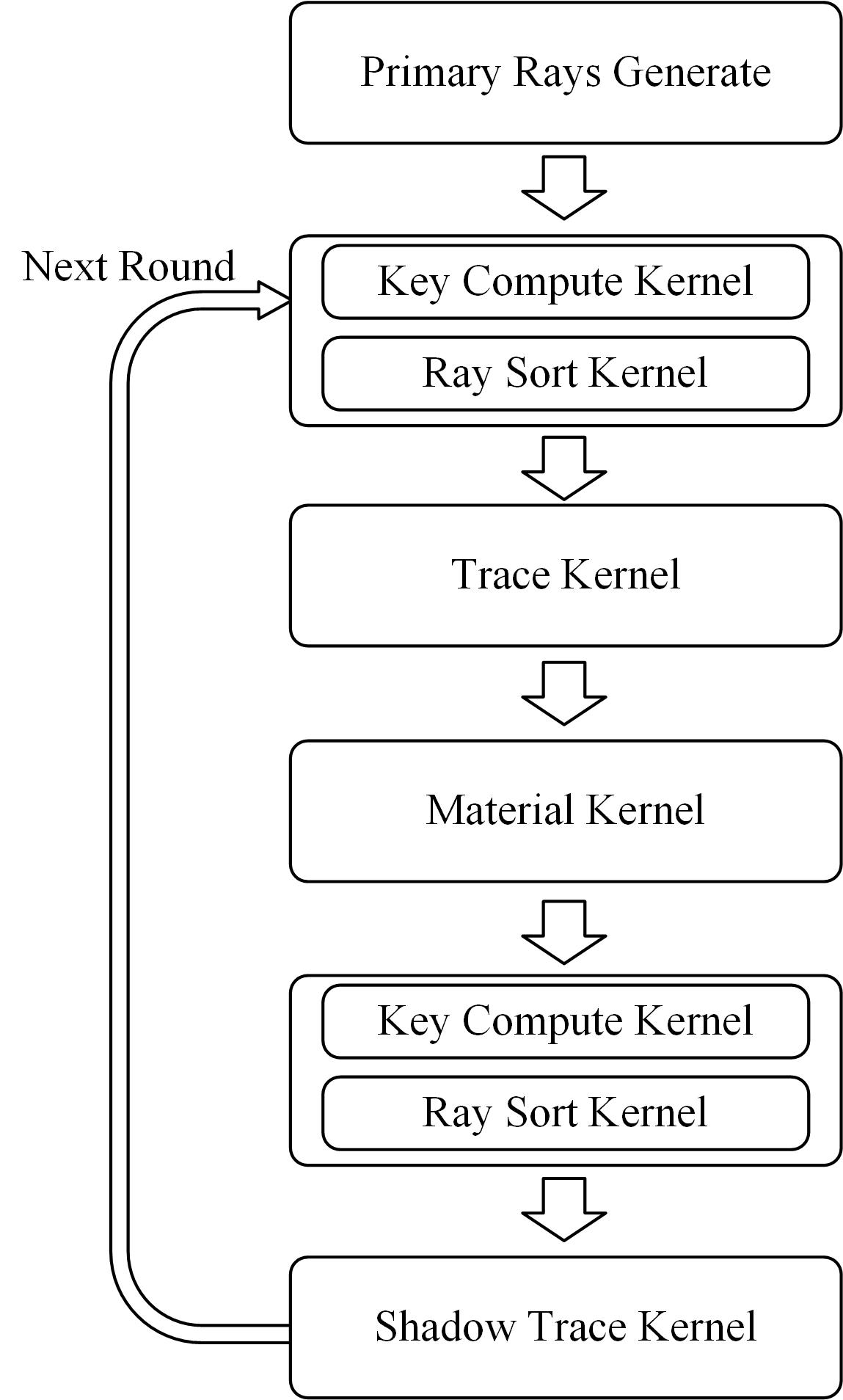}
\caption{Overall render pipeline diagram.}

\label{Fig_pipeline}
\end{figure}

\begin{figure*}[!htbp]
\centering
\subfigcapskip=3pt 
\subfigure[Breakfast scene.]{
\begin{minipage}{0.35\linewidth}
	\centering
	\includegraphics[width=\linewidth]{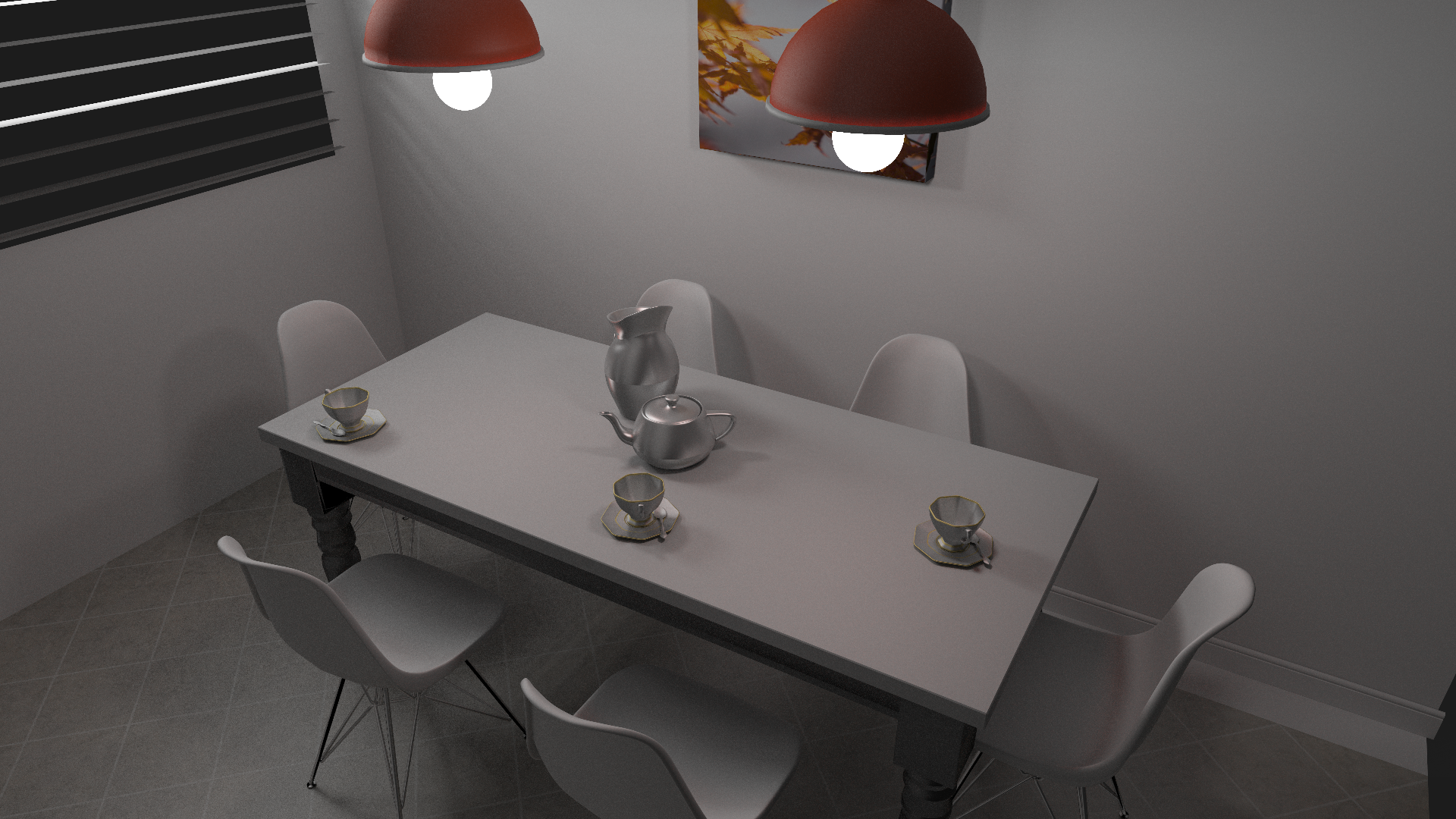}
\end{minipage}
}
\subfigure[Cut node boundaries in Breakfast scene.]{
\begin{minipage}{0.35\linewidth}
	\centering
	\includegraphics[width=\linewidth]{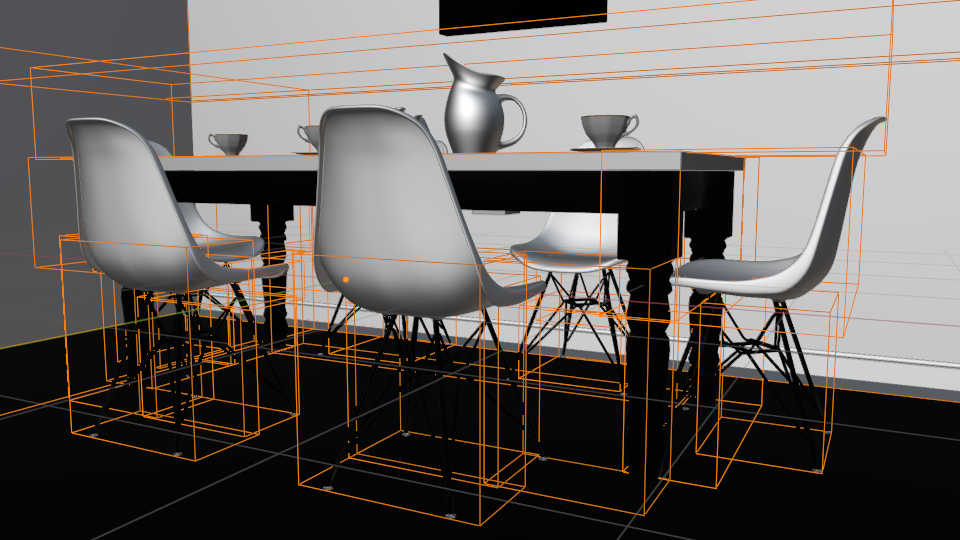}
\end{minipage}
}

\subfigure[Salle de Bain scene.]{
\begin{minipage}{0.35\linewidth}
	\centering
	\includegraphics[width=\linewidth]{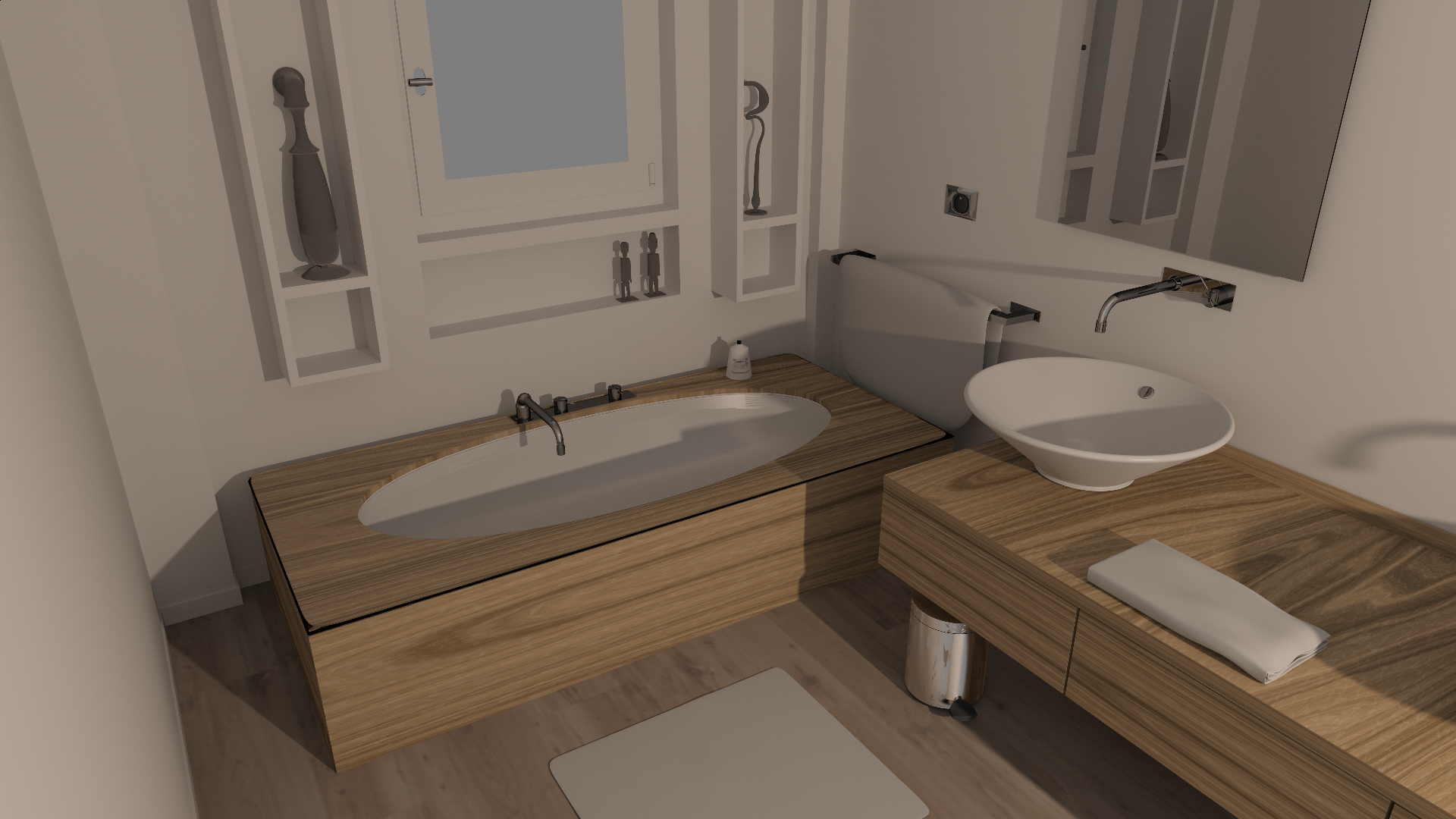}
\end{minipage}
}
\subfigure[Cut node boundaries in Salle de Bain scene.]{
\begin{minipage}{0.35\linewidth}
	\centering
	\includegraphics[width=\linewidth]{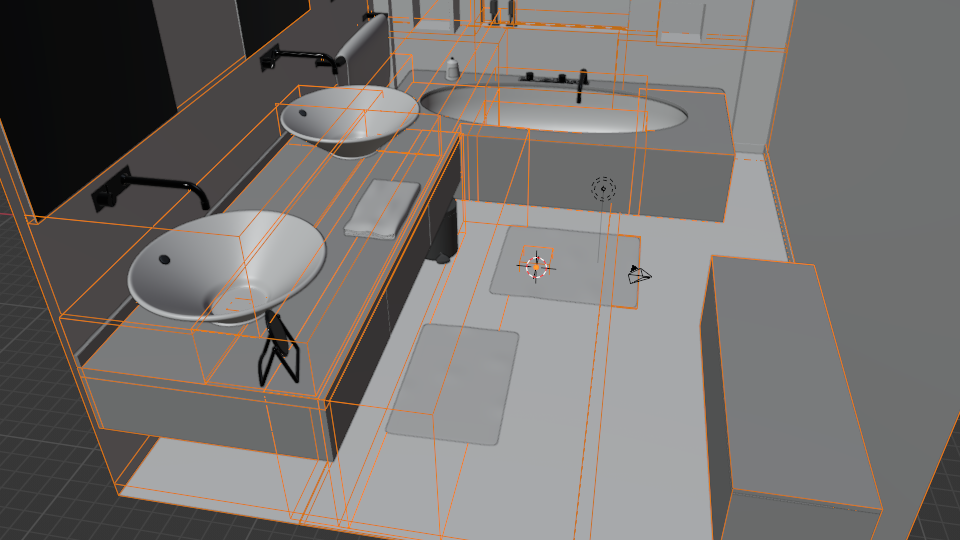}
\end{minipage}
}

\subfigure[Living room scene.]{
\begin{minipage}{0.35\linewidth}
	\centering
	\includegraphics[width=\linewidth]{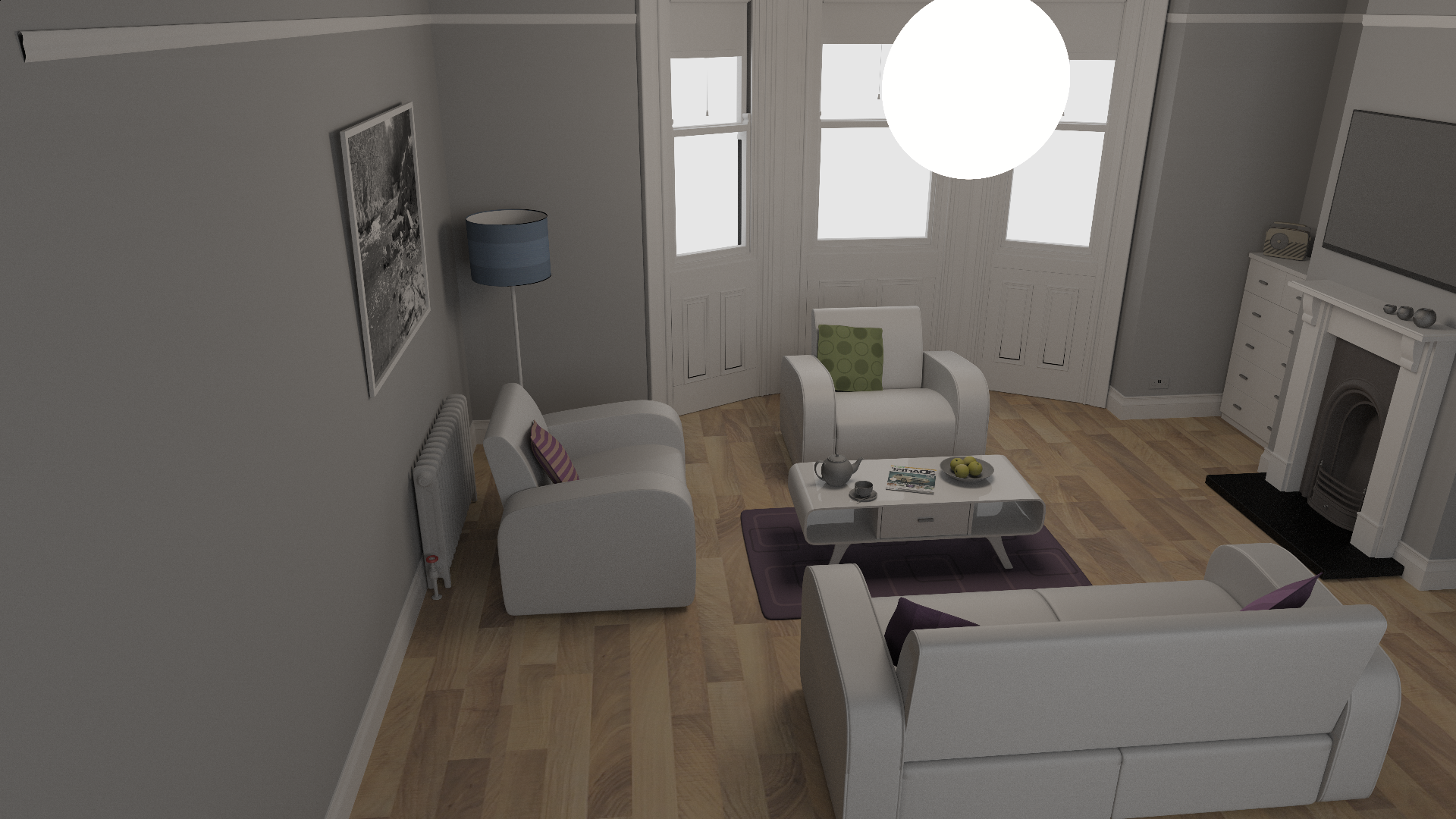}
\end{minipage}
}
\subfigure[Cut node boundaries in Living room scene.]{
\begin{minipage}{0.35\linewidth}
	\centering
	\includegraphics[width=\linewidth]{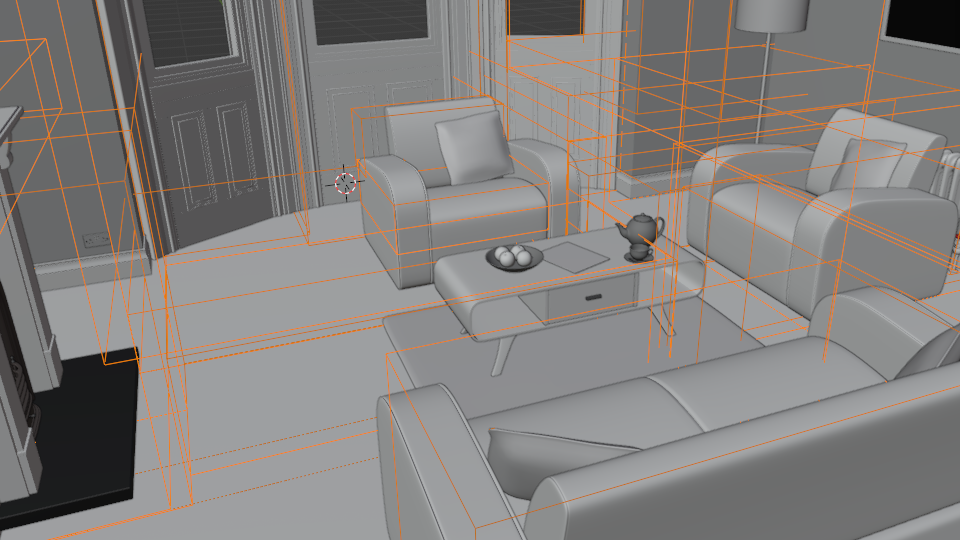}
\end{minipage}
}

\subfigure[Sibenik scene.]{
\begin{minipage}{0.35\linewidth}
	\centering
	\includegraphics[width=\linewidth]{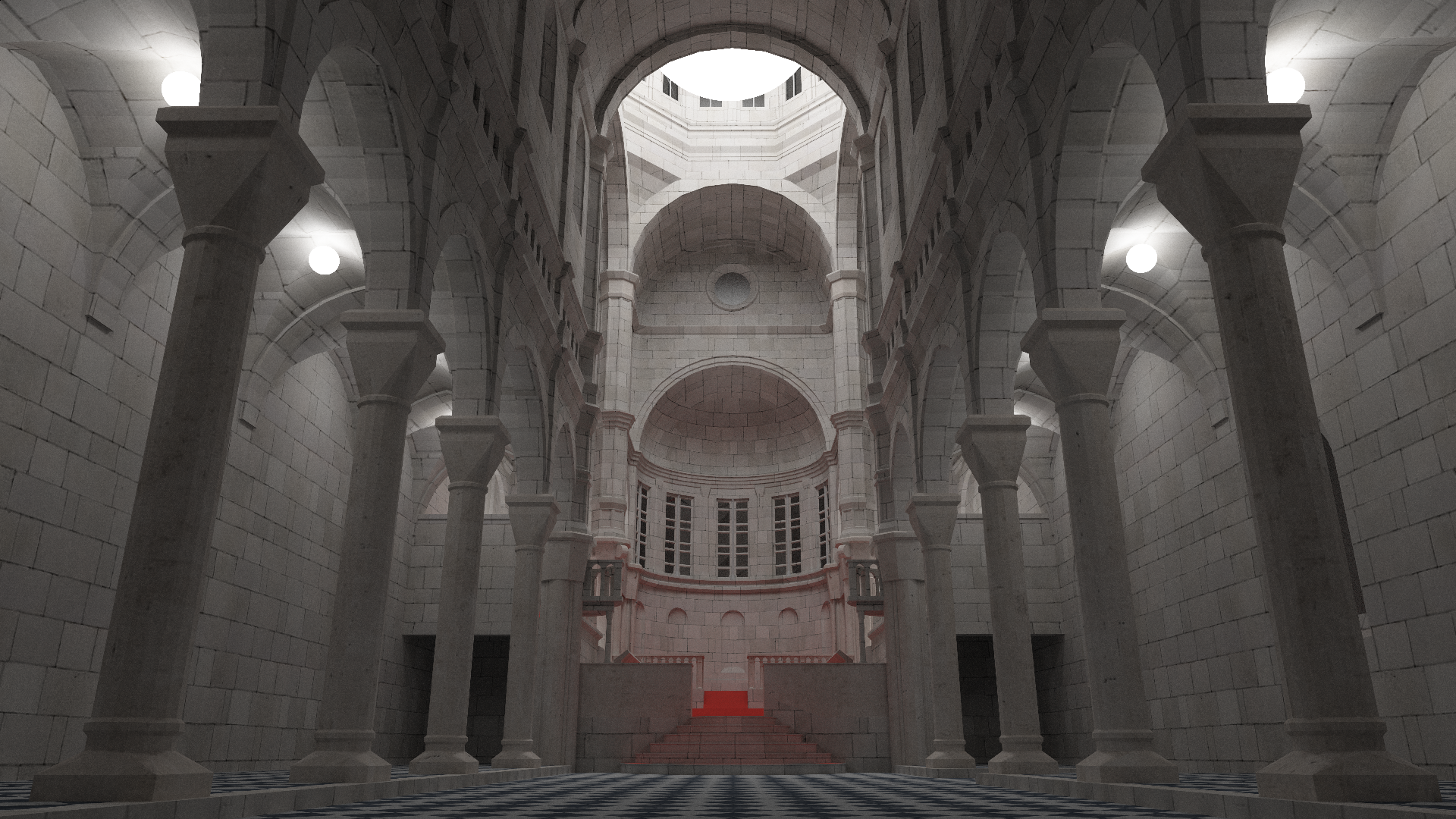}
\end{minipage}
}
\subfigure[Cut node boundaries in Sibenik scene.]{
\begin{minipage}{0.35\linewidth}
	\centering
	\includegraphics[width=\linewidth]{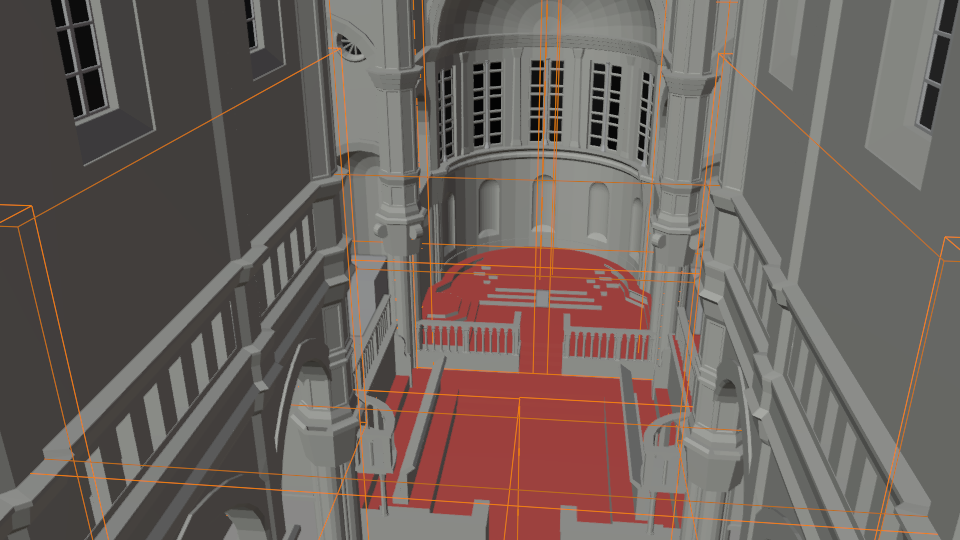}
\end{minipage}
}

\subfigure[Vokselia scene.]{
\begin{minipage}{0.35\linewidth}
	\centering
	\includegraphics[width=\linewidth]{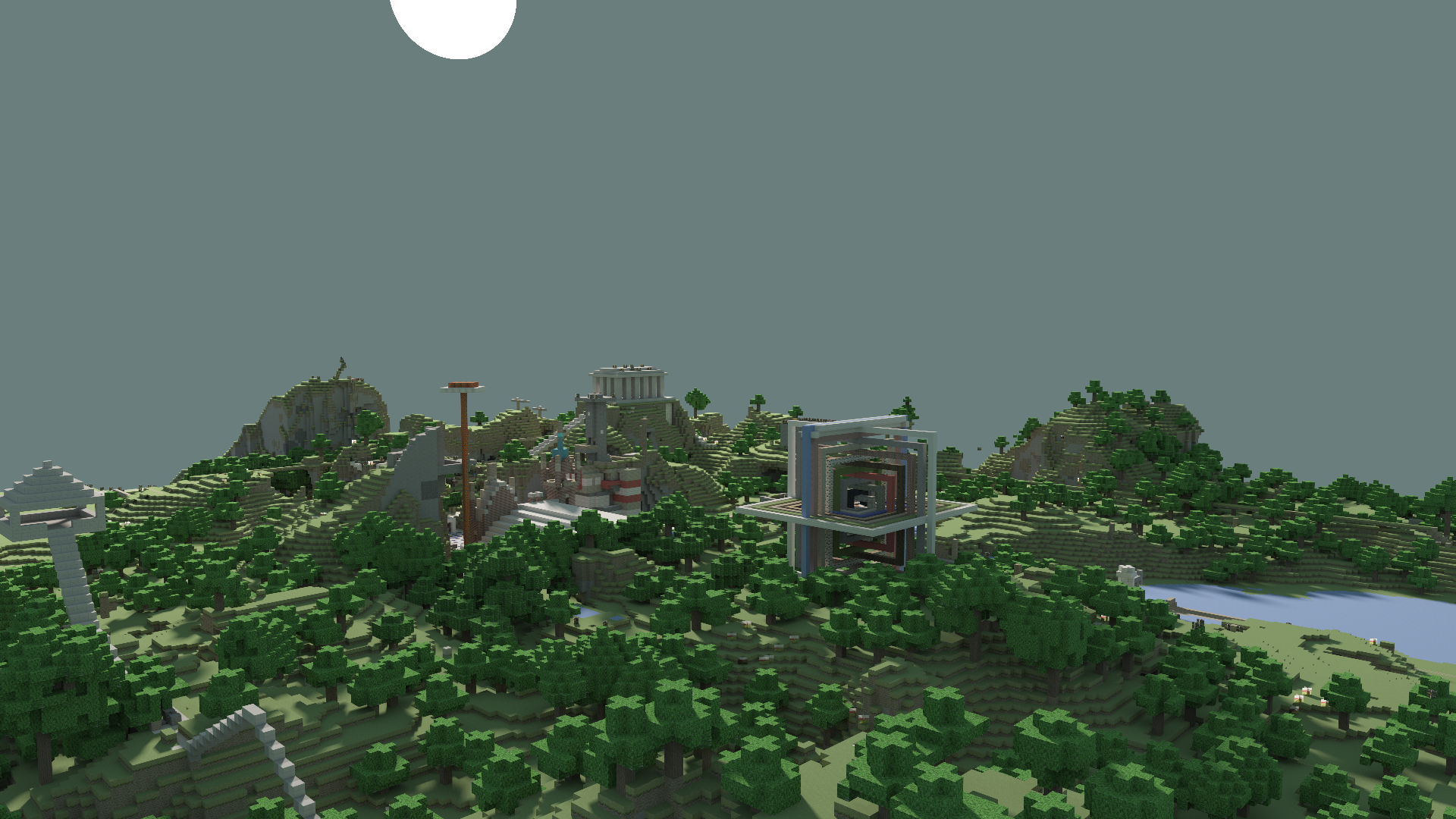}
\end{minipage}
}
\subfigure[Cut node boundaries in Vokselia scene.]{
\begin{minipage}{0.35\linewidth}
	\centering
	\includegraphics[width=\linewidth]{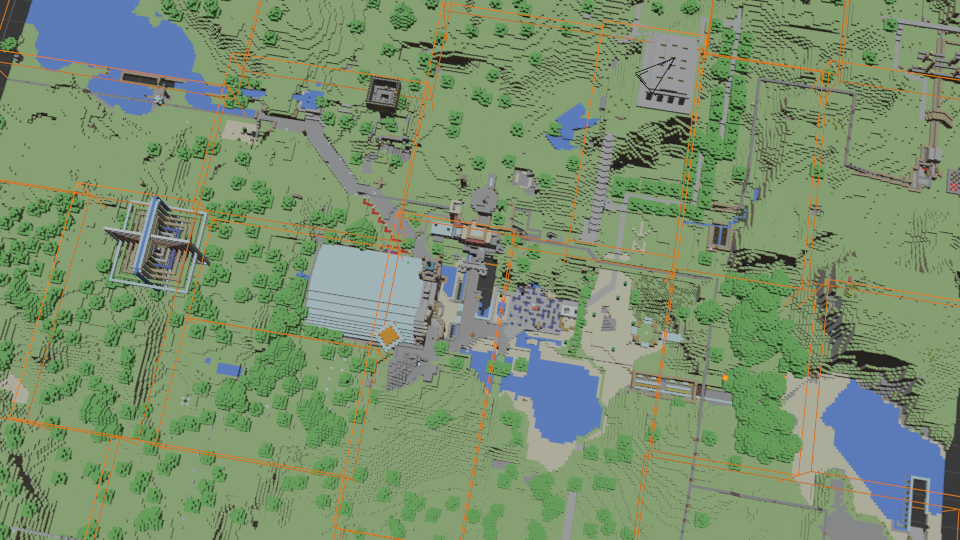}
\end{minipage}
}

\caption{Cut node boundaries (orange wireframe boxes) of hierarchy cut code in five scenes.}
\label{Fig_CutBoundariesSample}
\end{figure*}

We implemente our ray reordering technique in CUDA C on NVIDIA GeForce RTX 3060 GPU. The path tracer utilizes next event estimation with two shadow rays per bounce and eight samples per pixel. The overall pipeline, as illustrated in Figure \ref{Fig_pipeline}, follows the small kernel framework proposed by Laine et al. \cite{Laine2013}.

In our implementation, we add a key compute kernel and a ray sort kernel before the trace kernel to incorporate the ray reordering technique. The ray sorting kernel is implemented using the Thrust library, which efficiently sorts the indices of rays. Subsequently, the trace kernel fetches the rays based on these sorted indices, ensuring coherent ray traversal. By integrating our ray reordering technique into the existing pipeline, we can achieve improved efficiency and performance in ray tracing.

%------------------------------------------------------------------------

%------------------------------------------------------------------------

\section{Results}

\begin{table}[]
\centering
\caption{Light source setting table.}
\label{tab_LightSource}
\begin{tabular}{lcc}
\hline
              & Light Index              & Relative Radius \\ \hline
Breakfast     & 0                        & 8.30\%          \\
              & 1-2  & 1.60\%          \\
Salle de Bain & 0                        & 1.10\%          \\
Living room    & 0                        & 5\%             \\
Sibenik       & 0                        & 7.50\%          \\
              & 1-2  & 2.50\%          \\
              & 3-12 & 0.50\%          \\
Vokselia      & 0                        & 12\%      \\
\hline     
\end{tabular}
\end{table}

The performance is evaluated in five scenes of various complexity. Thanks to McGuire \cite{McGuire2017Data} and Bitterli \cite{resources16} for providing the necessary models. The light sources in the scenes include sky light and spherical light sources. The number and relative radius of the light sources are shown in Table \ref{tab_LightSource}. The first column of the table represents the index of the corresponding light source in the scene, and the second column represents the relative radius of that light source, which is expressed as a percentage of the largest extent of the scene. Some scenes in the table use sphere light sources of different radii. For example, the Breakfast scene has two types of spherical light sources. The radius of the large light source (light 0) accounts for 8.3\% of the largest extent of the scene, while the radius of the two small light sources (light 1 and light 2) accounts for 1.6\%. This simple light source setting allows us to perform experiments without increasing the complexity of the original scenes. The measurements were performed on a GeForce RTX 3060 GPU with an image resolution of 1920 x 1080. The code was compiled using CUDA 10.0. All recorded data represents averages. Each scene was run three times, and the data from 100 frames were collected to calculate the average running time and the number of rays. We set $\lambda$ = 0.125 for our method and average distance = 0.25 for TwoPoint method. To select the appropriate hyperparameters, we have carried out experiments in many scenes. The results are shown in appendix \ref{hyperparameters}.

%------------------------------------------------------------------------

%------------------------------------------------------------------------

%------------------------------------------------------------------------

%------------------------------------------------------------------------

\begin{table*}[]
\centering
\caption{Secondary ray trace speed comparison of our methods and test methods. The trace speed is measured in million rays per second (MRay/s). The numbers in parentheses represent the acceleration ratio of the method.}
\label{tab_Results_Overview}
\resizebox{\linewidth}{26mm}{
\begin{tabular}{lcccccccc}
 \hline
Scene         & Breakfast  & Salle de Bain & Living room  & Sibenik     & Vokselia & MultiInstance & San & landscape   \\ \hline
\#Triangles   &     1347K       &       1231K        &      580K       &     75K        &     1876K & 309K & 5.6M & 21.84M     \\ \hline
              &    
\begin{minipage}[b]{0.2\columnwidth}
\centering
\raisebox{-.5\height}{\includegraphics[width=\linewidth]{IMG/breakfast}}
\end{minipage}         &    
\begin{minipage}[b]{0.2\columnwidth}
\centering
\raisebox{-.5\height}{\includegraphics[width=\linewidth]{IMG/bathroom}}
\end{minipage}
           &         
\begin{minipage}[b]{0.2\columnwidth}
\centering
\raisebox{-.5\height}{\includegraphics[width=\linewidth]{IMG/Living}}
\end{minipage}
               &      
\begin{minipage}[b]{0.2\columnwidth}
\centering
\raisebox{-.5\height}{\includegraphics[width=\linewidth]{IMG/sib}}
\end{minipage}
                      &    
\begin{minipage}[b]{0.2\columnwidth}
\centering
\raisebox{-.5\height}{\includegraphics[width=\linewidth]{IMG/MC}}
\end{minipage}
 &    
\begin{minipage}[b]{0.2\columnwidth}
\centering
\raisebox{-.5\height}{\includegraphics[width=\linewidth]{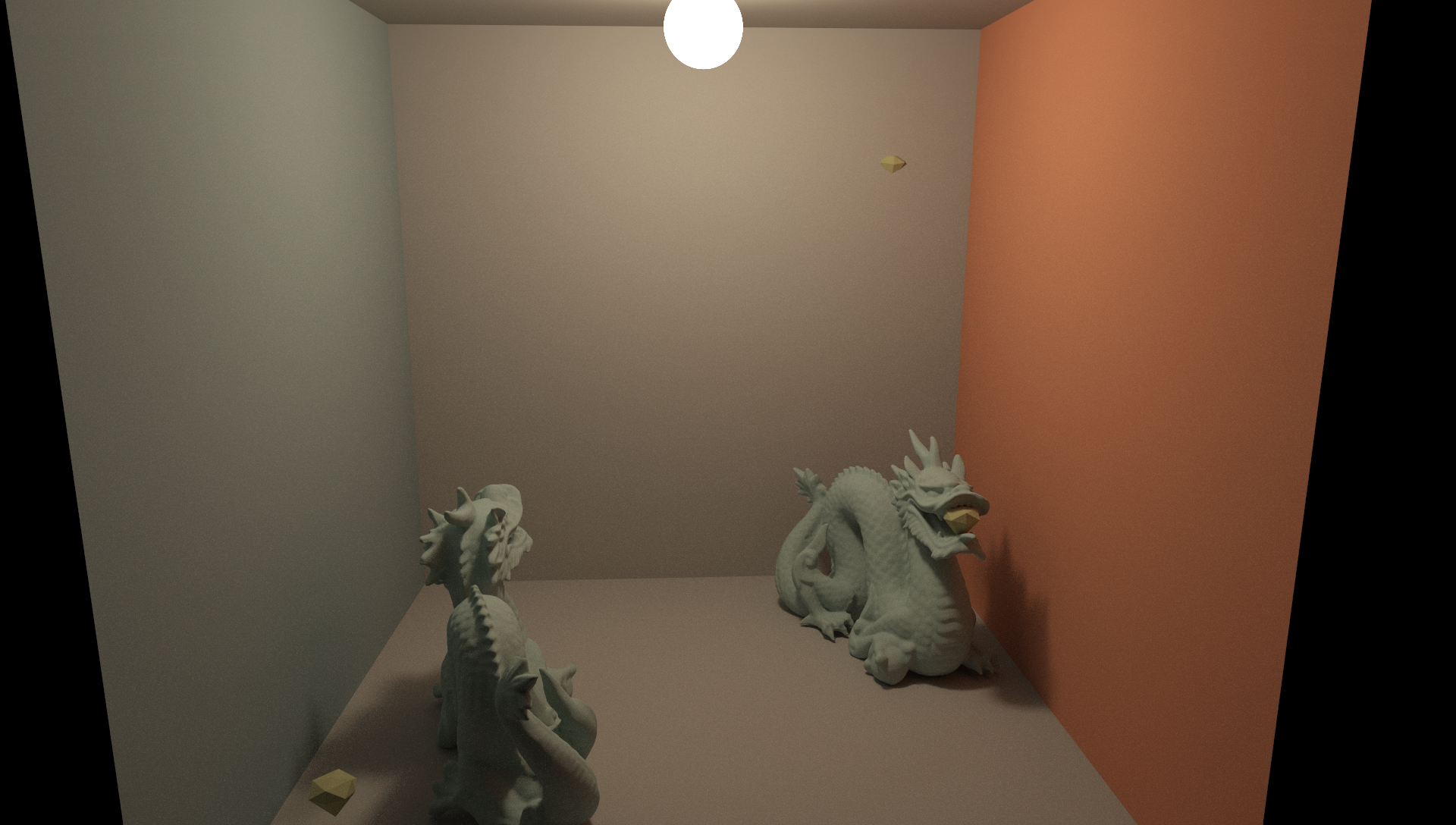}}
\end{minipage}
 &    
\begin{minipage}[b]{0.2\columnwidth}
\centering
\raisebox{-.5\height}{\includegraphics[width=\linewidth]{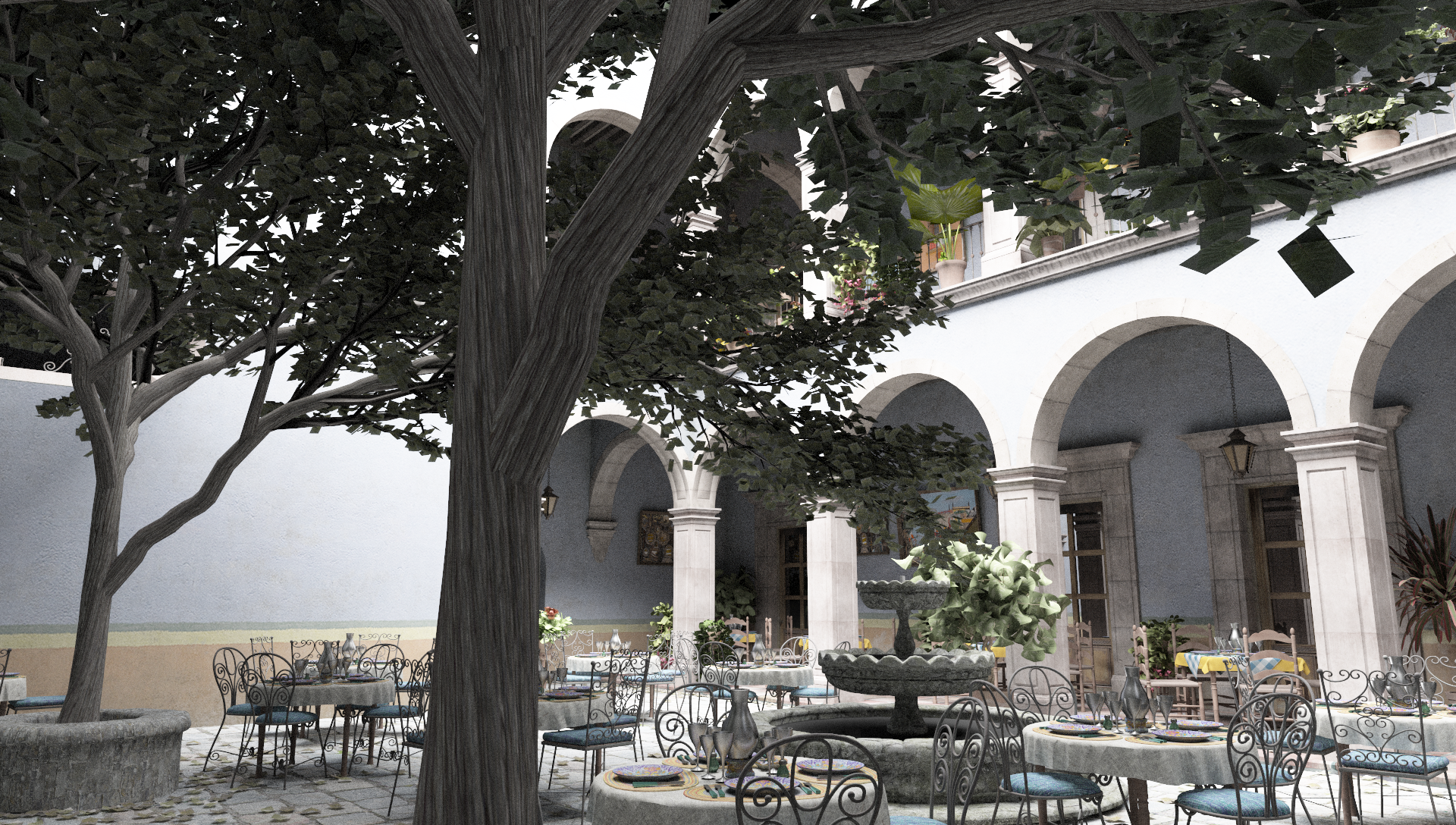}}
\end{minipage}
 &    
\begin{minipage}[b]{0.2\columnwidth}
\centering
\raisebox{-.5\height}{\includegraphics[width=\linewidth]{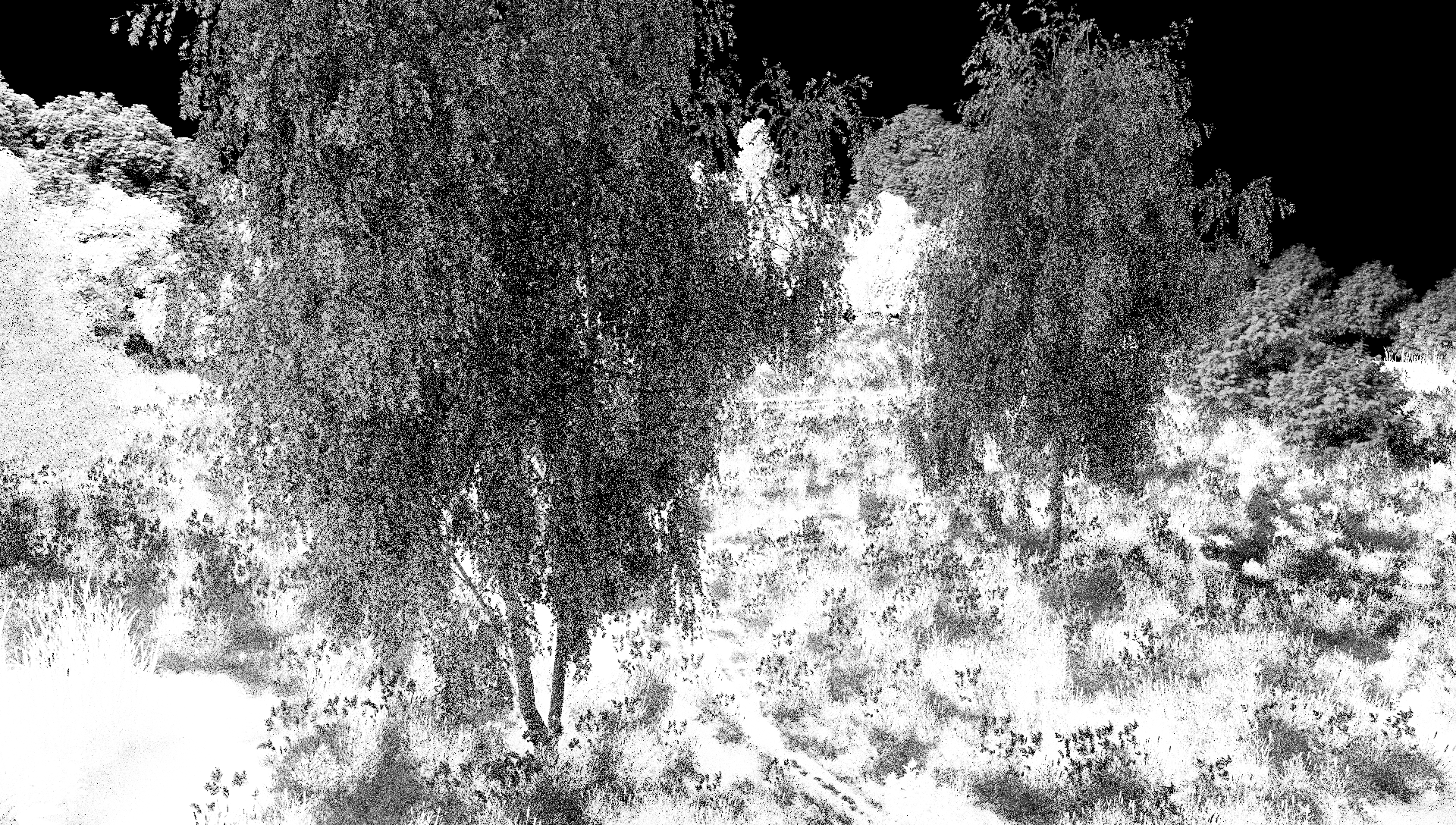}}
\end{minipage}
                              \\ \hline
NoSort      & 37.4 (1.00) & 206.1 (1.00)   & 189.4 (1.00) & 154.7 (1.00) & 64.2 (1.00) & 209.5(1.00)& 32.7(1.00)& 0.0464(1.00)\\
HCC (Ours)    & \textbf{67.6 (1.81)} & \textbf{284.9 (1.38)}   & 277.1 (1.46) & 208.1 (1.35) & 71.7 (1.12)& 270.9(1.29)& 41.6(1.26)& 0.0486(1.05) \\
MLHCC32 (Ours)  & 65.1 (1.74) & 280.2 (1.36)   & \textbf{277.8 (1.47)} & \textbf{218.5 (1.41)} & 74.4 (1.16)& \textbf{275.0(1.31)}& 39.8(1.21) & 0.0482(1.04)\\
MLHCC16 (Ours)  & 58.4 (1.56) & 254.0 (1.23)   & 250.6 (1.32) & 207.6 (1.34) & 73.2 (1.14)& 263.7(1.26)& 38.6(1.18)& 0.0492(1.06) \\
Regular Grid (Ours)  & 51.4 (1.37) & 247.1 (1.19)   & 243.3 (1.28) & 188.7 (1.22) & 70.7 (1.10)& 216.4(1.03)& 38.4(1.17)& 0.0440(0.95) \\
TwoPoint    & 56.8 (1.52) & 274.2 (1.33)   & 272.1 (1.44) & 217.9 (1.41) & 75.5 (1.18)& 242.0(1.15)& 39.5(1.21)& 0.0437(0.94) \\
Aila\_Compact & 55.6 (1.48) & 268.0 (1.30)   & 258.2 (1.36) & 208.9 (1.35) & \textbf{75.5 (1.18)}& 195.7(0.93)& 38.7(1.18)& 0.0465(1.00) \\
Reis          & 49.7 (1.33) & 242.4 (1.18)   & 224.8 (1.19) & 183.2 (1.18) & 74.3 (1.16)& 202.2(0.97)& \textbf{37.0(1.31)}& 0.0474(1.02) \\
Origin        & 46.1 (1.23) & 233.0 (1.13)   & 208.3 (1.10) & 165.4 (1.07) & 74.8 (1.17)& 199.0(0.95)& 35.1(1.08)& \textbf{0.0501(1.08)} \\ \hline
PPC           & 68.3 (1.82) & 342.0 (1.66)   & 348.5 (1.84) & 278.0 (1.80) & 87.1 (1.36)& 211.0(1.01)& 41.3(1.26)& 0.0491(1.06) \\ \hline

\end{tabular}}
\end{table*}

\subsection{Ray Tracing Performance Overall} 

In three scenes, our methods do not achieve the best performance with a 32-bit length. To analyze this phenomenon, we provide the cut node boundaries of the hierarchy cut code in Figure \ref{Fig_CutBoundariesSample}. In the Breakfast, Salle de Bain, and Living room scenes, the cut nodes match the details in the scene, which are the most complex parts of the scenes. Allocating more bits to such details can effectively encode the rays. However, the experiments on the Vokselia and landscape scenes reveal that reordering techniques face challenges in dealing with open landscape scenes. All the reordering techniques only achieve a speedup of 1.1 or 1.2 in these scenes. The reason why our methods fail in these scenes is that there are too many details contained within one code boundary. For instance, the forest in Vokselia is present in only one cut node, which cannot accurately determine whether a ray passes through this forest. Similar reasons explain the limitations of other existing encoding methods. For encoding methods designed for the entire scene, open landscape scenes pose a significant challenge. It is worth considering dividing the scene into different regions according to the viewport and encoding rays in different regions separately.

In the Sibenik, Vokselia, and MultiInstance scenes, there are large empty areas while many details gather in small areas. For example, most of the details in Sibenik are concentrated on the wall at the edge of the scene. Therefore, the HCC can only roughly represent the boundary of the scene, but not the internal space. This causes most of the rays to pass through only 3-4 cut nodes, further exacerbating the problem of insufficient cut nodes. On the other hand, MLHCC can dynamically select a top-level cut node and split it into several smaller nodes, which can represent more valuable details. Hence, MLHCC performs better in these scenes. However, the 32-bit MLHCC does not achieve better results than the HCC in most small scenes. As analyzed in the last paragraph of Section \ref{MLHCCs}, MLHCC discards much information about top-level cut nodes, which may lead to less effective coding after compression.

In the MultiInstance scene, we set several dragon and diamond instances in a box. The results show that object instancing impacts the PPC performance because PPC wastes lots of bits in encoding high-level nodes of these instances, resulting in an inability to represent the details of the scenes. On the other hand, MLHCC still performs well because it uses two-level cut nodes and the low-level cut nodes vary according to the rays. This means that the cut nodes of MLHCC contain deeper nodes of BVH, which represent more details. However, if a scene has too many instances and each one is complex, such as landscape scene, the speedup of MLHCC will be reduced as well. In fact, nearly all encoding methods do not work well in this type of scene because there are too many details contained within one code boundary, as we have explained in the first paragraph of this section.

The speedup improvement achieved by our methods comes from avoiding boundary drift, rather than the advantage of using the intersection result of the bounding box and ray to encode the ray. The fifth row of Table \ref{tab_Results_Overview} shows the trace speed using regular grid encoding. As mentioned in Section 5, regular grid encoding is essentially the same as hierarchy cut code, except that it does not use BVH cut. From the data in Table \ref{tab_Results_Overview}, regular grid encoding does not offer any performance advantages compared to hierarchy cut code or the TwoPoint method because it uses a regular grid as the coding boundary, which will encounter the boundary drift problem. This confirms the necessity of eliminating boundary drift.

\begin{table*}[]
\centering
\caption{Primary shadow ray trace  comparison of our methods and test methods. The trace speed is measured in million rays per second (MRay/s). The numbers in parentheses represent the acceleration ratio of the method.}
\resizebox{\linewidth}{23mm}{
\label{tab_Results_Shadow_Overview}
\begin{tabular}{lcccccccc}
 \hline
Scene           & Breakfast & Salle de Bain & Living room & Sibenik & Vokselia  & MultiInstance & San & landscape \\ \hline
\#Triangles     & 1347K & 1231K & 580K & 75K & 1876K & 309K & 5.6M & 21.84M  \\ \hline
NoSort          & 152.4 (1.00) & 700.7 (1.00) & 425.6 (1.00) & 163.5 (1.00) & 501.3 (1.00) & 749.7 (1.00) & 53.9 (1.00) & 0.0649 (1.0000) \\
HCC (Ours)      & 230.6 (1.51) & 850.8 (1.21) & 522.9 (1.23) & 253.6 (1.55) & 537.5 (1.07) & 839.2 (1.12) & 78.0 (1.45) & \textbf{0.0698 (1.0756)} \\
MLHCC32 (Ours)  & 238.5 (1.56) & 821.0 (1.17) & 507.9 (1.19) & 234.6 (1.44) & 521.1 (1.04) & \textbf{845.8 (1.13)} & 72.3 (1.34) & 0.0655 (1.0086) \\
MLHCC16 (Ours)  & 206.1 (1.35) & 795.2 (1.13) & 499.3 (1.17) & 245.7 (1.50) & 520.4 (1.04) & 790.1 (1.05) & 74.3 (1.38) & 0.0686 (1.0569) \\
Regular Grid (Ours)    & 156.4 (1.03) & 688.9 (0.98) & 419.8 (0.99) & 201.9 (1.29) & 491.6 (0.98) & 574.8 (0.77) & 43.3 (0.80) & 0.0585 (0.9010) \\
TwoPoint      & 257.1 (1.69) & 975.4 (1.39) & 561.0 (1.32) & 314.5 (1.92) & 521.7 (1.04) & 710.0 (0.95) & \textbf{86.7 (1.60)} & 0.0605 (0.9324) \\
Aila\_Compact   & 275.7 (1.81) & 1032.1 (1.47) & \textbf{580.0 (1.36)} & \textbf{324.1 (1.98)} & 632.0 (1.26) & 621.1 (0.83) & 84.1 (1.56) & 0.0627 (0.9658) \\
Reis            & \textbf{276.6 (1.81)} & 1048.4 (1.50) & 577.1 (1.36) & 268.2 (1.64) & 664.5 (1.33) & 644.3 (0.86) & 84.2 (1.56) & 0.0645 (0.9932) \\
Origin          & 245.6 (1.61) & \textbf{1090.9 (1.56)} & 527.9 (1.24) & 204.8 (1.25) & \textbf{747.0 (1.49)} & 649.4 (0.87) & 75.0 (1.39) & 0.0655 (1.0096) \\ \hline
PPC             & 287.8 (1.89) & 1208.9 (1.73) & 682.1 (1.60) & 323.0 (1.98) & 820.7 (1.64) & 726.5 (0.97) & 79.8 (1.48) & 0.0651 (1.0022) \\ \hline

\end{tabular}}
\end{table*}

Reordering techniques can accelerate arbitrary reflection rays, not just secondary rays. The discussion regarding this topic can be found in Appendix \ref{reflection}.

\subsection{Shadow Ray Acceleration}

The reordering method can also improve the tracing efficiency of shadow rays, as indicated in Table \ref{tab_Results_Shadow_Overview}. We evaluated the performance of primary shadow ray tracing. However, when it comes to accelerating shadow rays, our method is not as efficient as conventional reordering techniques. While the TwoPoint method achieves a speedup of 1.92x in the Sibenik scene, the HCC only achieves a speedup of 1.55x in the same scene. This discrepancy is attributed to the characteristics of shadow rays. Since shadow rays always point towards the light source, encoding the direction or origin of the rays can be more effective in such cases, especially when there is only one small light source in the scene. Conversely, if the HAD value of the cut node containing the sole light source is high, the bit corresponding to this cut node will be nearly wasted because it will be the same for all the HCCs of shadow rays. The HCC achieves relatively good speedup results in the Breakfast and Sibenik scenes. These scenes have multiple light sources, as indicated in Table \ref{tab_LightSource}, and thus the shadow rays are not all directed towards a single location.

\begin{figure}[!htbp]
\centering
\subfigcapskip=3pt 
\subfigure[Breakfast scene in limited viewport.]{
\begin{minipage}{0.45\linewidth}
	\centering
	\includegraphics[width=\linewidth]{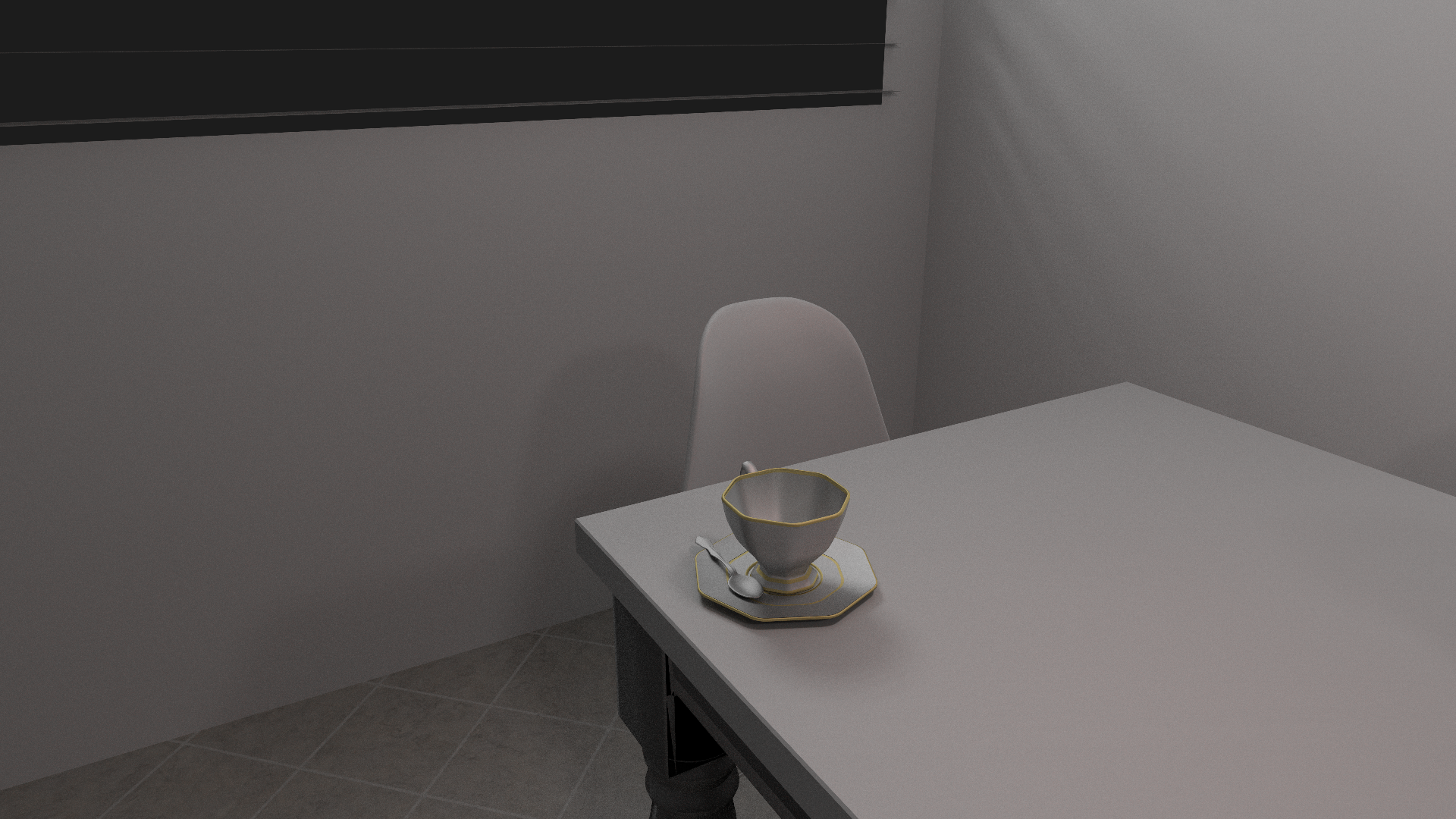}
\end{minipage}
}
\subfigure[Salle de Bain scene in limited viewport.]{
\begin{minipage}{0.45\linewidth}
	\centering
	\includegraphics[width=\linewidth]{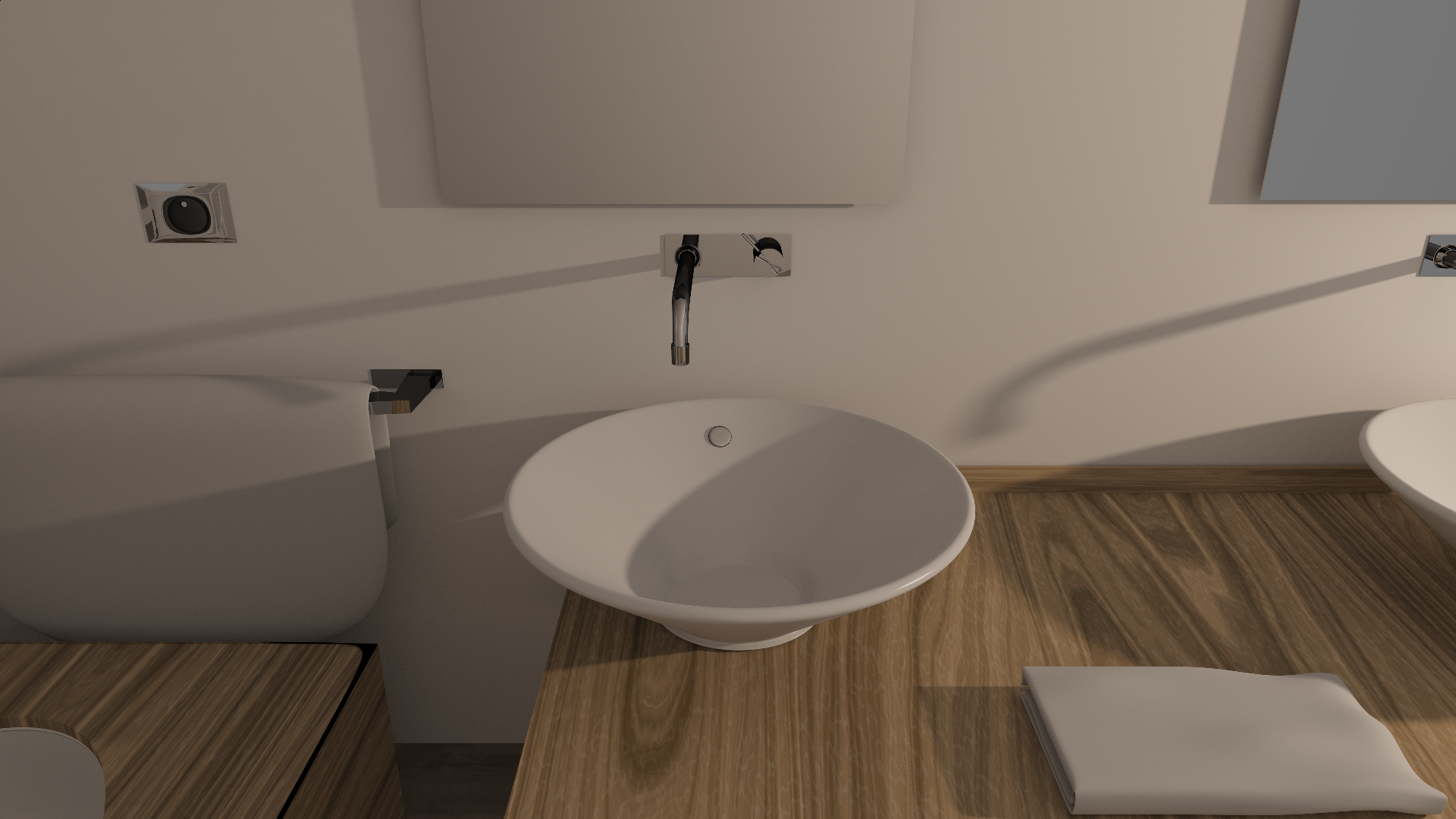}
\end{minipage}
}

\subfigure[Living room scene in limited viewport.]{
\begin{minipage}{0.45\linewidth}
	\centering
	\includegraphics[width=\linewidth]{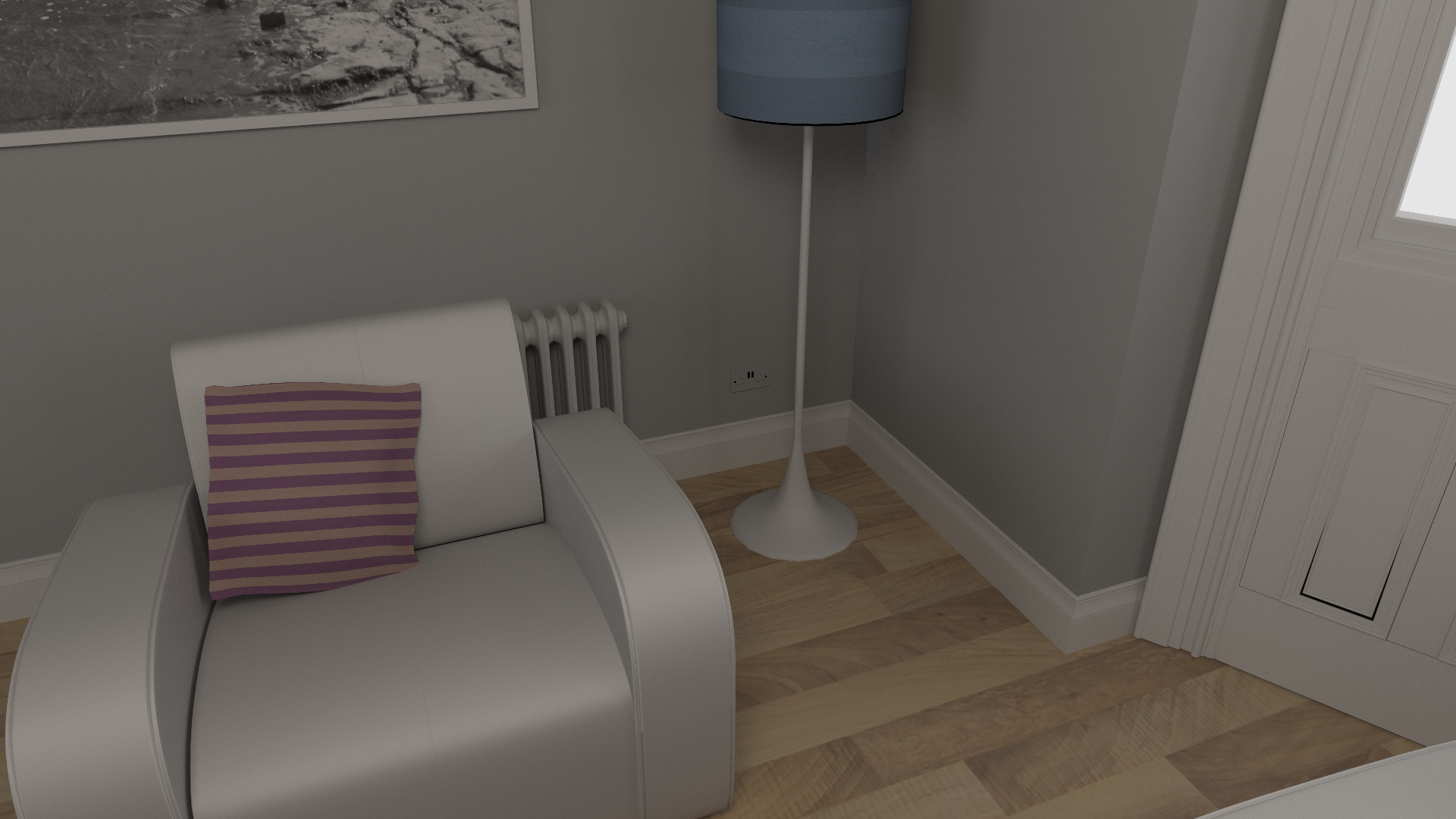}
\end{minipage}
}
\subfigure[Sibenik scene in limited viewport.]{
\begin{minipage}{0.45\linewidth}
	\centering
	\includegraphics[width=\linewidth]{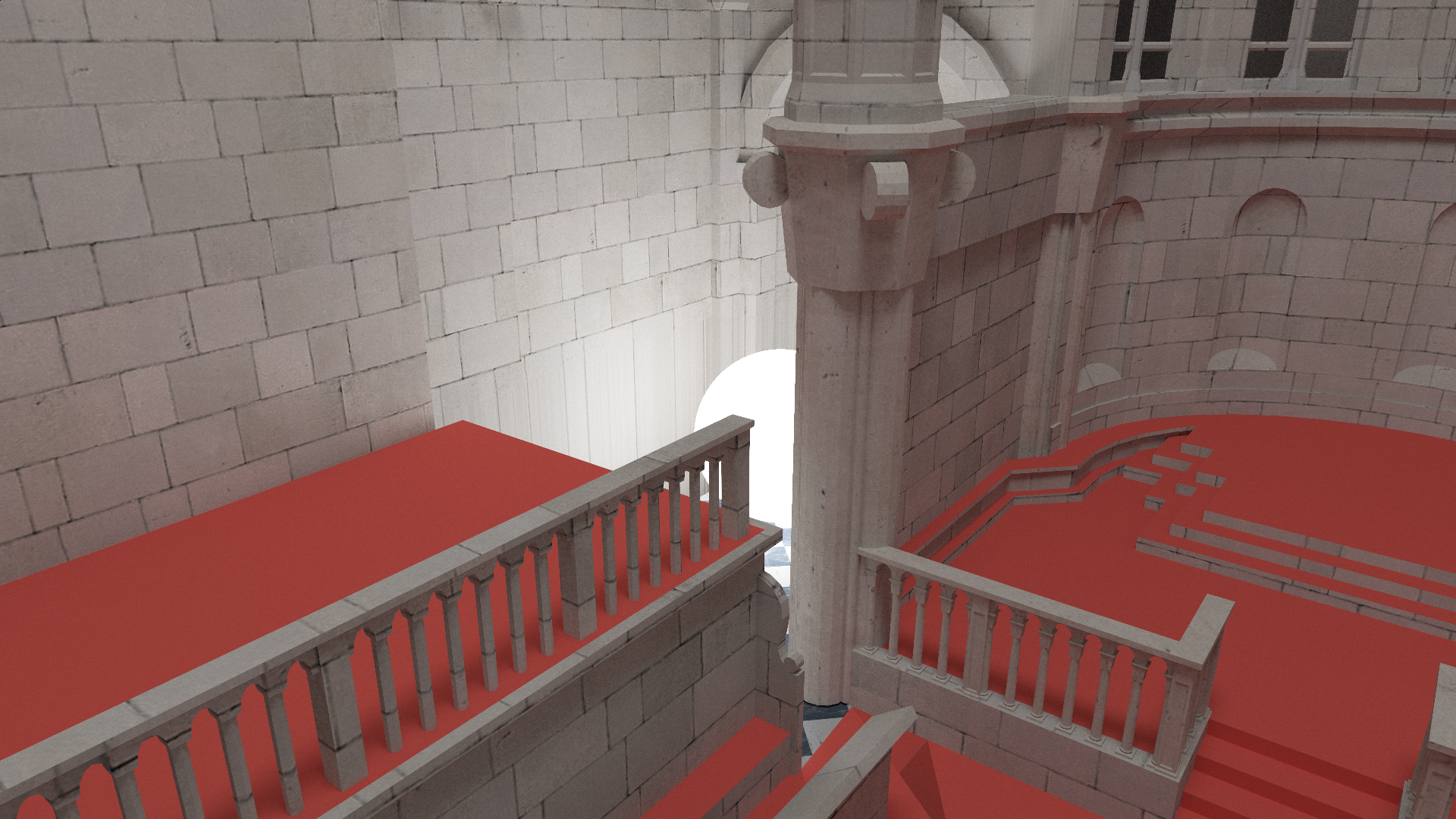}
\end{minipage}
}
\caption{Four scenes rendered in limited viewport. The speedup ratio of hierarchy cut code in these scenes are: (a) 1.81 in Breakfast, (b) 1.42 in Salle de Bain, (c) 1.49 in Living room, (d) 1.49 in Sibenik.}
\label{Fig_LV}
\end{figure}

\subsection{Performance in Limited Viewport}

\begin{table*}[]
\centering
\caption{Secondary ray trace speed comparison of our methods and test methods in limited vierport. The trace speed is measured in million rays per second (MRay/s). The numbers in parentheses represent the acceleration ratio of the method.}
\label{tab_limited_viewport}
\begin{tabular}{lcccc}
\hline
Scene     & Breakfast  & Salle de Bain    & Living room  & Sibenik         \\
\hline
&    
\begin{minipage}[b]{0.3\columnwidth}
\centering
\raisebox{-.5\height}{\includegraphics[width=\linewidth]{IMG/LVbreakfast}}
\end{minipage}        
&      
\begin{minipage}[b]{0.3\columnwidth}
\centering
\raisebox{-.5\height}{\includegraphics[width=\linewidth]{IMG/LVbathroom}}
\end{minipage}         &
\begin{minipage}[b]{0.3\columnwidth}
\centering
\raisebox{-.5\height}{\includegraphics[width=\linewidth]{IMG/LVlivingroom}}
\end{minipage}               &
\begin{minipage}[b]{0.3\columnwidth}
\centering
\raisebox{-.5\height}{\includegraphics[width=\linewidth]{IMG/LVSik}}
\end{minipage}               \\
\hline
NoSort    & 41.1(1.00) & 170.7(1.00) & 132.5(1.00) & 125.2(1.00) \\
HCC (Ours) & 74.4(1.81) & 241.3(1.41) & 197.5(1.49) & 186.9(1.49) \\
MLHCC32 (Ours)   & \textbf{77.9(1.90)} & 235.1(1.38) & 188.3(1.42) & 194.9(1.56) \\
MLHCC16 (Ours)   & 73.1(1.78) & 220.6(1.29) & 165.4(1.25) & 183.9(1.47) \\
Regular Grid (Ours)       & 62.4(1.52) & 217.1(1.27) & 178.5(1.35) & 166.4(1.33) \\
TwoPoint  & 63.8(1.55) & \textbf{243.4(1.43)} & \textbf{202.8(1.53)} & \textbf{199.9(1.60)} \\
Aila\_Compact      & 61.2(1.49) & 236.7(1.39) & 193.3(1.46) & 188.6(1.51) \\
Reis      & 52.9(1.29) & 217.8(1.28) & 169.4(1.28) & 161.8(1.29) \\
Origin    & 44.7(1.09) & 204.6(1.20) & 148.0(1.12) & 133.8(1.07) \\ \hline
PPC       & 83.8(2.04) & 281.0(1.65) & 249.0(1.88) & 236.1(1.88) \\ \hline
\end{tabular}
\end{table*}

The experiment on shadow ray acceleration demonstrates that the distribution of rays can impact the effectiveness of the hierarchy cut code. Additionally, a limited viewport, where only a small portion of the scene is visible on the screen, can also affect the ray distribution. In such cases, rays tend to hit the same small area in the scene and then scatter into the larger scene. Figure \ref{Fig_LV} illustrates rendered images with a limited viewport, and Table \ref{tab_limited_viewport} provides an overview of the performance in these limited viewport scenarios. By comparing Table \ref{tab_limited_viewport} with Table \ref{tab_Results_Overview}, we can conclude that our method exhibits sufficient robustness to handle the impact of limited viewports effectively.
 
In the Salle de Bain, Living room, and Sibenik scenes, our method performs well but does not outperform the TwoPoint method. The reason for this result is similar to what we explained in the shadow ray acceleration part. In the case of secondary rays, the limited viewport causes them to have similar origins but different endpoints. Therefore, encoding the ray's direction or endpoint can be more effective in such scenarios. 

%------------------------------------------------------------------------

%------------------------------------------------------------------------

\subsection{Overhead of Reordering Techniques}

Evaluating the overhead is an important aspect of assessing the performance of reordering techniques. It allows us to understand the additional computational cost incurred by the encoding and sorting processes. Table \ref{tab_overhead} presents the overhead of each method in the Breakfast scene. While Reis, Aila\_Compact, and TwoPoint methods have the same overhead, we only display the overhead for the TwoPoint method. The percentage column represents the ratio of the total overhead (including encoding and sorting overhead) to the tracing time. The results indicate that our HCC has a similar overhead to the ordinary methods. Additionally, the MLHCC (16 bits) exhibits lower sorting and encoding overhead because it uses a shorter sorting key. This demonstrates the feasibility of using compression coding to reduce overhead. 

\begin{table}[]
\caption{Overhead of reordering techniques in Breakfast scene.}
\label{tab_overhead}
\begin{tabular}{lcccc}
\hline
           & \makecell[c]{Encode \\ (ms)}  & \makecell[c]{Sort\\ (ms)}  & \makecell[c]{Sum\\ (ms)} & Percentage \\ \hline
HCC (Ours)        & 4.51                  & 9.35   &  13.86            & 5.66\%              \\
MLHCC16 (Ours)    & 3.36                  & 5.47  &        8.83       & 3.19\%              \\
MLHCC32 (Ours)   & 5.76                   & 9.38  &   15.14           & 5.63\% \\
TwoPoint & 2.73                  & 9.44   &          12.17    & 4.24\%              \\
PPC        & 24.93                 & 9.41    &          34.34   & 13.06\%             \\ \hline   
\end{tabular}
\end{table}

\subsection{Reordering Techniques with Longer Sorting Keys}

Table \ref{tab_differentBits} presents the performance of secondary ray tracing in the Sibenik scene using sorting keys of different bit lengths. The patterns of Origin, Reis, Aila\_Compact, and TwoPoint methods are the same when using different lengths of sorting keys, we only display the pattern of the TwoPoint method. Interestingly, our methods (PPC, HCC, MLHCC) show a performance improvement with longer sorting keys, whereas ordinary methods do not exhibit this trend. This suggests that our methods can effectively utilize longer sorting keys to enhance the tracing efficiency.

\begin{table}[]
\centering
\caption{The trace speed (MRays/s) with different sorting key bit length in Sibenik scene.}
\label{tab_differentBits}
\begin{tabular}{llll}
\hline
                      & 16bits        & 32bits        & 64bits      \\ \hline
HCC (Ours)                   & 187.2         & 208.1         & 222.0       \\
MLHCC (Ours)                & 207.6         & 218.5         & 240.9       \\
TwoPoint            & 193.5         & 217.9         & 220.5       \\
PPC                   & 213.9         & 274.7         & 293.7       \\ \hline
\end{tabular}
\end{table}

Longer sorting keys can make ray encoding more accurate and result in a finer coding boundary. However, it is important to note that the ordering of rays within the same warp does not affect the tracing efficiency. Even with a finer encoding, the order of rays that already have different codes will not change. For example, if two rays are encoded as 1 and 2, their order will remain the same even if they receive a finer coding with longer sorting keys (e.g., 1.x and 2.y, where x and y represent any decimal number). The benefit of longer sorting keys comes into play when multiple warps have the same code, indicating a rough encoding. In such cases, longer sorting keys can change the order of rays and improve the overall tracing efficiency.

\begin{table}[]
\centering
\caption{Average number of codes with different sorting key bit length in Sibenik scene.}
\label{tab_ANC}
\begin{tabular}{llll}
\hline
                      & 16bits        & 32bits        & 64bits      \\ \hline
HCC (Ours)            & 1.02          & 1.17          & 1.73        \\
MLHCC (Ours)                & 1.06          & 1.52          & 2.87        \\
TwoPoint            & 1.88          & 13.94         & 13.94       \\
PPC                   & 1.05          & 4.33          & 18.02       \\ \hline
\end{tabular}
\end{table}

To further validate our analysis, we calculated the average number of codes (ANC) for each encoding method. The ANC represents the average number of different codes obtained in one warp when rays are encoded using a specific method. The results are presented in Table \ref{tab_ANC}. Notably, the TwoPoint method exhibits a significantly higher ANC than the other methods when using a 32-bit sorting key. The ANC reflects the adequacy of the encoding. 

Analyzing Table \ref{tab_ANC}, we observe that all methods have a small ANC when using a 16-bit sorting key. This indicates that their encoding is insufficient at this level. Increasing the length of the sorting keys can enhance tracing efficiency. However, the TwoPoint method already exhibits a relatively high ANC with a 32-bit sorting key. This suggests that the rays have been fully encoded, and extending the sorting key length has minimal impact on tracing efficiency. This explains why the TwoPoint method with a 64-bit sorting key does not achieve significant performance improvement compared to the 32-bit version. As mentioned in Section 3.2, the TwoPoint method is prone to boundary drift. Thus, when the encoding is sufficient, its tracing efficiency is inferior to that of the hierarchy cut code. On the other hand, the other three methods exhibit small ANC values with a 32-bit sorting key, indicating that they can still benefit from longer sorting keys.

In conclusion, increasing encoding accuracy does not always translate into improved tracing efficiency. When an encoding method has a high ANC value, the encoding is already sufficient. In such cases, extending the sorting key length does not provide significant performance gains. Therefore, increasing coding accuracy is not a universal solution. Instead, a crucial direction for improvement is ensuring the consistency between traversal path coherence and ray spatial coherence as measured by the encoding method.

%------------------------------------------------------------------------
\subsection{Discussion}
\subsubsection{Application in Faster Tracing Kernels}

The reordering method needs to achieve higher speedup and lower overhead to be effectively applied in faster tracking cores. Meister et al.'s work highlighted the challenge of large overhead, which often leads to low speedup or even negative speedup on RTX platforms. In this paper, we propose two feasible approaches to reduce the overhead. One is to employ compression schemes like MLHCC to improve efficiency. Since the arrangement of rays within the same warp does not impact the tracing efficiency, sorting these rays may not be necessary. Taking measures to eliminate unnecessary work can enhance the sorting speed. For instance, we can skip the last few bits during encoding and terminate bucket sorting early. Alternatively, heuristic methods could be explored to trade-off sorting accuracy for faster sorting speed. Hardware acceleration schemes may also be viable options.

It is worth noting that our encoding process, which involves accessing the acceleration structure, cannot be directly applied to the RTX core within the existing framework. Consequently, we did not conduct experiments on the RTX core. However, our work addresses two key challenges identified in Meister et al.'s research regarding reordering techniques, namely the large overhead and bit bottleneck. While our work does not provide the final answer or solution to the question of whether the ray reordering preprocessing step is beneficial on the current RTX platform, it does contribute to answering this question. If the reordering method can indeed accelerate the tracing process, it holds the potential for improved implementation efficiency through hardware-based approaches.

\subsubsection{Better Encoding Method}
For reordering technology, it is important to note that a better encoding scheme does not necessarily mean a more accurate encoding scheme. The analysis in Section 6.5 highlights that simply increasing the length of the sorting key does not guarantee improved tracing performance. Longer sorting keys can only enhance the tracing performance when the encoding itself is still insufficient, as demonstrated by our hierarchy cut code. However, it is worth exploring other avenues to address this challenge beyond longer sorting keys.

One potential approach is to investigate better cut nodes searching methods or compression techniques.  Another possibility is to combine our encoding method with traditional encoding techniques, which may offer a solution to address both the issue of insufficient encoding and boundary drift simultaneously. It is important to note that this approach may potentially increase the encoding time, but the benefits in terms of improved performance may outweigh the cost.

\subsubsection{Adapting Large Scene}
The experiments have demonstrated that existing reordering techniques struggle to achieve significant acceleration results in large-scale scenes. In such scenarios, each ray only utilizes a small portion of the code, resulting in a considerable waste of encoding bits that represent parts of the scene the ray will never access. To address this issue, it becomes necessary to employ scene partitioning methods to ensure that the encoding bits are not wasted on areas that are not within the current perspective.

%------------------------------------------------------------------------

%------------------------------------------------------------------------

\section{Conclusion}

In this paper, we introduce the hierarchy cut code and multi-level hierarchy cut code as solutions to mitigate the negative impact of boundary drift on ray reordering techniques. By encoding rays based on the cut of BVH, we effectively address the performance issues caused by boundary drift. The hierarchical accumulation of density within the code enables it to automatically adapt to areas with dense meshes, resulting in improved tracing performance and higher robustness. Furthermore, our multi-level hierarchy cut code achieves similar performance to other methods while using only half the sorting key length. This highlights the efficiency and effectiveness of our encoding scheme in reducing the encoding overhead. Additionally, we analyze and tackle the issue faced by ordinary reordering techniques, which fail to benefit significantly from longer sorting keys. Our results demonstrate that targeting the encoding of the acceleration structure, regardless of the spatial position, can yield more versatile and effective reordering effects. This finding provides important guidance for the further development of ray reordering techniques in the field.

%------------------------------------------------------------------------

%------------------------------------------------------------------------
\section{Acknowledgments}
This work was supported by National Key R\&D Program of China (No: 2021YFF0500302) and National Natural Science Fund of China (No: 61502185). We would like to thank Benedikt Bitterli and Morgan McGuire for providing the test scenes.

%------------------------------------------------------------------------

% bibtex
%\bibliographystyle{eg-alpha-doi}  
%\bibliography{my-ref}        

% biblatex with biber
\printbibliography                

%-------------------------------------------------------------------------
\appendix
\section {Hierarchy Cut Code with Direction Encoding}

Encoding the direction information into hierarchy cut code will not improve the performance. From Table \ref{tab_HCC_DIR}, we can see that the hierarchy cut code with direction information (HCC\_DIR) performance even worse than 31 bit version. That is because the direction information introduces boundary drift to the code.

\begin{table}[h]
\centering
\caption{The trace speed of  hierarchy cut code with direction information.}
\label{tab_HCC_DIR}
\begin{tabular}{lccc}
\hline
              & \multicolumn{3}{c}{Trace Speed (MRays/s)} \\ \hline
              & HCC       & HCC\_DIR     & HCC\_31bit     \\ \hline
Breakfast     & 67.6      & 60.6         & 65.3           \\
Salle de Bain & 284.9     & 278.8        & 284.1          \\
Living room    & 277.1     & 263.1        & 275.9          \\
Sibenik       & 208.1     & 203          & 203.5          \\
Vokselia      & 71.7      & 71.4         & 71.1          \\ \hline
\end{tabular}
\end{table}

\section {Hyperparameters Selection}\label{hyperparameters}
\begin{figure}[htb]
\centering
\includegraphics[width=\linewidth]{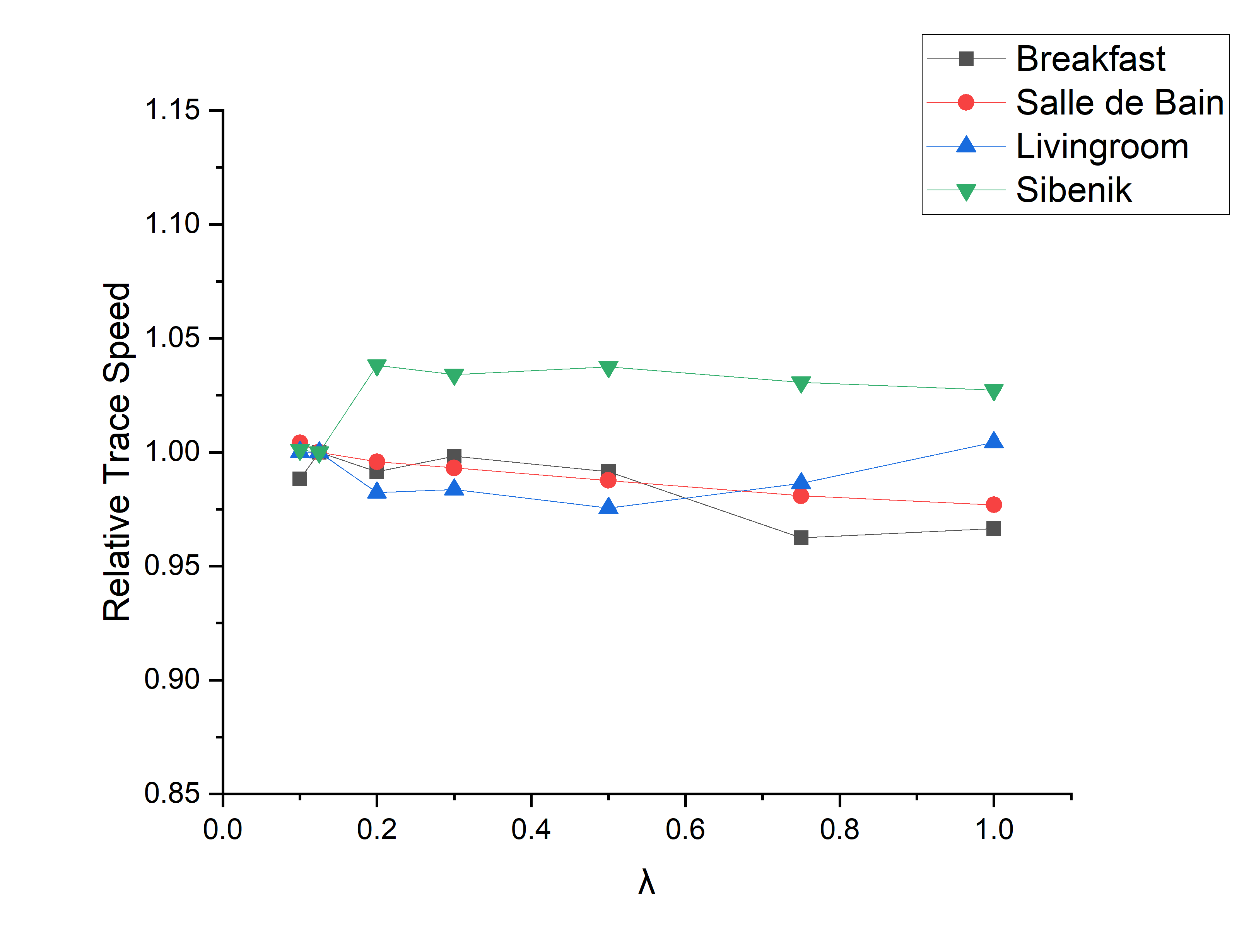}
\caption{The trace performance of hierarchy cut code with different $\lambda$ value.}

\label{Fig_lambda}
\end{figure}
\begin{figure}[htb]
\centering
\includegraphics[width=\linewidth]{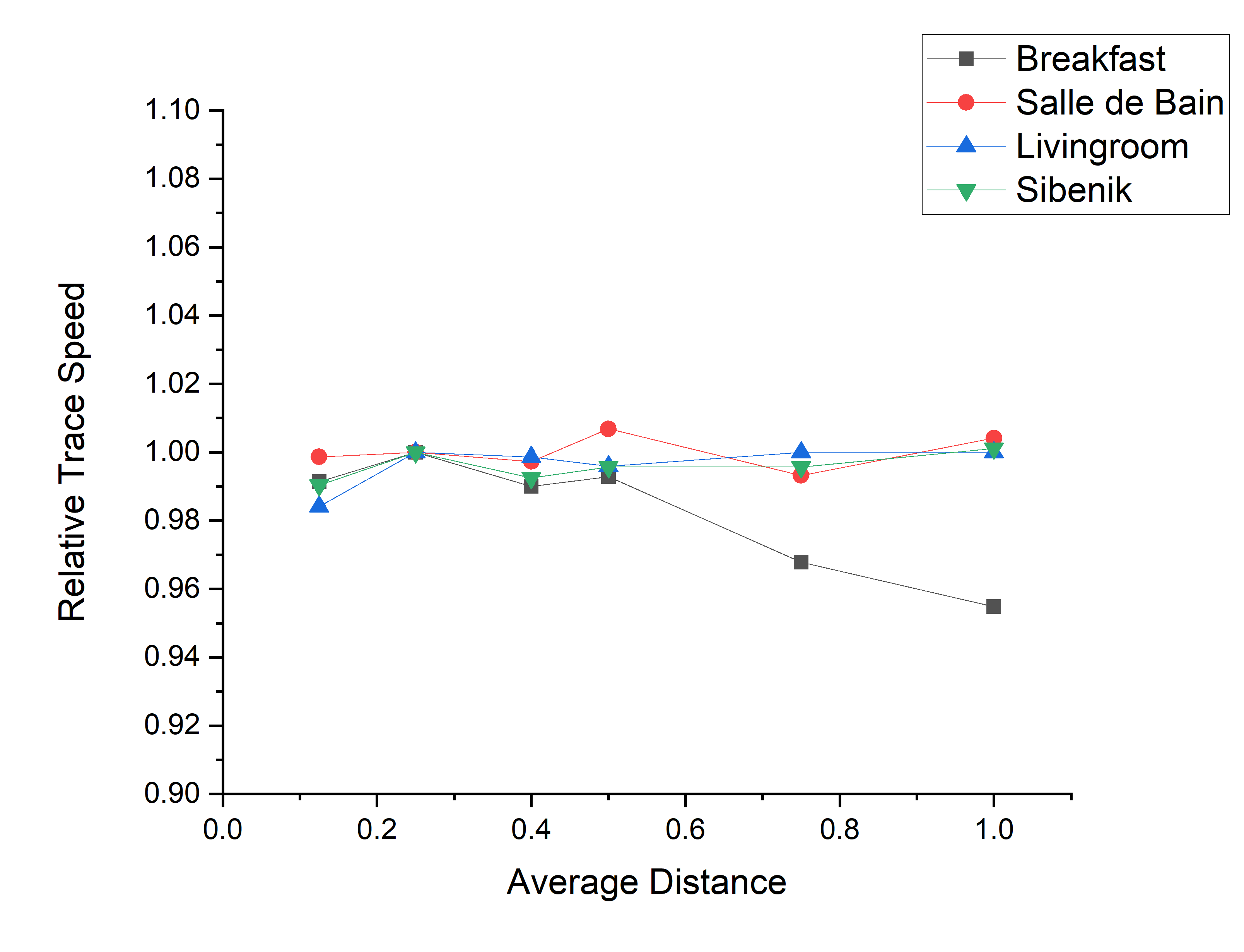}
\caption{The trace performance of TwoPoint method with different average distance.}

\label{Fig_avgdis}
\end{figure}

The hyperparameter $\lambda$ will affect the weight of internal node surface area when calculating the HAD value. We choose some parameter values unevenly between 0.1-1 and plot the trace speed in Figure \ref{Fig_lambda}. Experiments have proved that $\lambda$ value of 0.125 is appropriate. Although for some scenes, this is not an optimal choice. In order to avoid the unfairness of the performance test caused by the over-adaptation scenes by hyperparameter selection, we does not use the best hyperparameter each scene. The result also shows that not every different $\lambda$ value will greatly affect the trace speed. This is because the  $\lambda$ affects the trace speed by affecting the selection of cut nodes. The breadth-first based search order determines that the upper nodes will be searched sooner or later. Changes in lambda values sometimes only affect one or two cut nodes. Therefore, the variation of $\lambda$ in 0.1 to 1 will not have a significant impact on trace speed.

The average distance will affect the termination point estimation for TwoPoint method. Meister et al.’ work used the value 0.25 to perform the test. We choose some parameter values unevenly between 0.1-1 and plot the trace speed in Figure \ref{Fig_avgdis}. The experiment shows that 0.25 is an appropriate choice.
Although the average distance is used to estimate the termination point, we believe it acts more like the mixed weight of origin and direction. The greater the average distance, the greater the influene of direction on estimated termination point. Therefore, the coding will contain more direction information.

\section {Reflection Ray Trace Performance}\label{reflection}
\begin{figure}[htb]
\centering
\includegraphics[width=\linewidth]{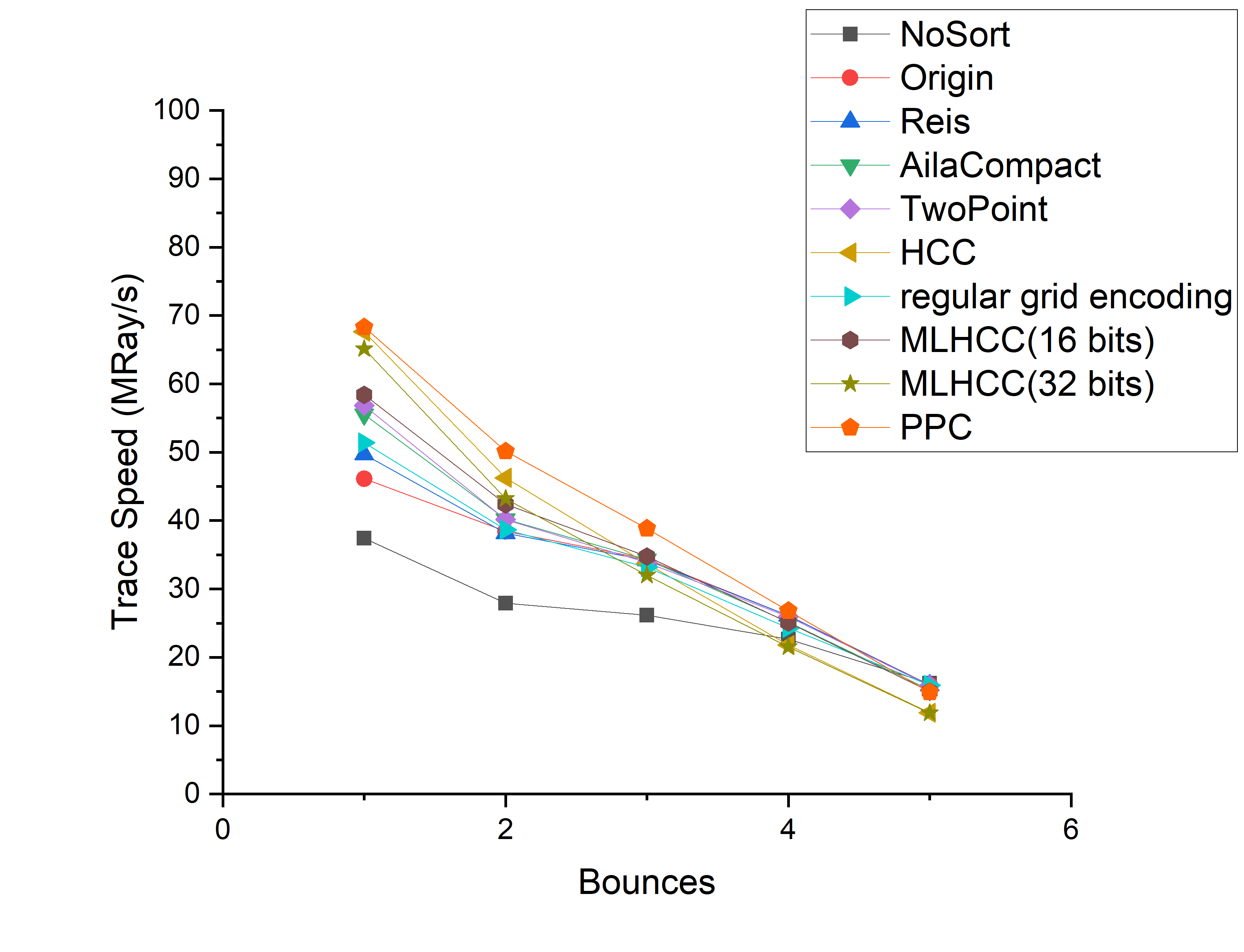}
\caption{The reflection ray trace performance in Breakfast scene for subsequent bounces.}

\label{Fig_performanceBounds}
\end{figure}

Figure \ref{Fig_performanceBounds} illustrates the impact of reordering on different reflection bounces in the breakfast scene. Each reflection bounce shows a certain level of acceleration achieved by the reordering technique. However, the speed of ray tracing still decreases as the number of reflection bounces increases, even with the application of the reordering technology. Deeper reflections exhibit a less pronounced speedup effect for two main reasons. Firstly, the use of Russian Roulette strategy in light reflection results in a decrease in the number of rays as the reflections progress leading to a less noticeable acceleration effect of the sorting method. Secondly, during the random reflection process, rays become more scattered, resulting in sparse and distant rays that still need to be traced at the final stages. Even with sorting, it becomes challenging to find rays with high traversal path consistency in such scenarios.
\end{document}